\def\lesssim{\mathrel{\hbox{\rlap{\hbox{\lower4pt\hbox{$\sim$}}}\hbox{$<$}}}}

\def\gtrsim{\mathrel{\hbox{\rlap{\hbox{\lower4pt\hbox{$\sim$}}}\hbox{$>$}}}}

\def\msun{M$_{\odot}$}

\def\ll_lsun{Log$({L/\rm L_{\odot}})$~}

\def\masa_msun{$M/ \rm M_{\odot}$~}

\def\m_mstar{$M/M_{*}$~}

\documentclass{aa}
\usepackage{graphicx}
        
\begin{document}

\title{The mode trapping properties of full DA  white dwarf 
evolutionary models}

\author{A. H. C\'orsico,\thanks{Fellow of the Consejo Nacional de 
Investigaciones Cient\'{\i}ficas y T\'ecnicas (CONICET), Argentina.}
L. G. Althaus,\thanks{Member of the Carrera del Investigador
Cient\'{\i}fico y Tecnol\'ogico, CONICET, Argentina.} 
O. G. Benvenuto\thanks{Member  of  the  Carrera  del Investigador
Cient\'{\i}fico, CIC, Argentina.}
\and A. M. Serenelli \thanks{Fellow of CONICET, Argentina.} }

\offprints{A. H. C\'orsico}

\institute{Facultad  de  Ciencias
Astron\'omicas  y Geof\'{\i}sicas, Universidad  Nacional de  La Plata,
Paseo  del  Bosque S/N,  (1900)  La  Plata,  Argentina\\ Instituto  de
Astrof\'{\i}sica de La Plata, IALP, CONICET\\
\email{acorsico,althaus,obenvenu,serenell@fcaglp.fcaglp.unlp.edu.ar} }

\date{Received; accepted}

\abstract{An adiabatic,  non-radial pulsation  study of a  0.563 \msun  
DA white  dwarf model  is presented on  the basis of  new evolutionary
calculations performed  in a self-consistent way  with the predictions
of time  dependent element diffusion, nuclear burning  and the history
of the white dwarf progenitor.   Emphasis is placed on the role played
by the  internal chemical  stratification of these  new models  in the
behaviour  of  the  eigenmodes,  and  the expectations  for  the  full
g-spectrum  of  periods.   The  implications  for  the  mode  trapping
properties are discussed at length.  In this regard, we find that, for
high  periods, the  viability of  mode  trapping as  a mode  selection
mechanism  is markedly  weaker for  our models,  as compared  with the
situation in  which the  hydrogen-helium transition region  is treated
assuming equilibrium diffusion in the trace element approximation.
 }  

\authorrunning{C\'orsico et al.}

\titlerunning{The adiabatic pulsation properties of full DA  white dwarf 
evolutionary models}

\maketitle


\section{Introduction} \label{sec:intro}

Since photometric variations  were detected in the white  dwarf HL Tau
76   (Landolt  1968),  astronomers   have  been   observing  multimode
pulsations in  an increasing number  of these objects.   Of particular
interest are the variable  white dwarfs characterized by hydrogen-rich
atmospheres.  These variable stars, known in the literature as ZZ Ceti
or DAV  stars, constitute the  most numerous group  amongst degenerate
pulsators.  Other  class of pulsating  white dwarfs are the  DBV, with
helium-rich  atmospheres, and  the  pre-white dwarfs  DOVs and  PNNVs,
which  show  spectroscopically pronounced  carbon,  oxygen and  helium
features  (for reviews of  the topic,  see Winget  1988 and  Kepler \&
Bradley  1995). In particular,  ZZ Ceti  stars are  found in  a narrow
interval of effective temperature ($T_{\rm eff}$) ranging from 12500 K
$\gtrsim T_{\rm  eff} \gtrsim$ 10700 K.   Their brightness variations,
which reach up to 0.30  magnitudes, are interpreted as being caused by
spheroidal, non-radial g(gravity)-modes of  low degree ($\ell \leq 2$)
and low  and intermediate  overtones $k$ (the  number of zeros  in the
radial eigenfunction), with periods  ($P_k$) between 2 and 20 minutes.
Radial modes,  although found overstables  in a number  of theoretical
studies  of pulsating  DA white  dwarfs  (see, e.g.   Saio et  al. 
  1983),  have   been  discarded  as  the  cause  of
variability in  such stars.  This  is so because the  periods involved
are  shorter than  10 seconds.   Observationally  these high-frequency
signatures  have not  been  detected  thus far.   With  regard to  the
mechanism that drives pulsations, the $\kappa-\gamma$ mechanism is the
traditionally accepted one  (Dolez \& Vauclair 1981 and  Winget et al.
1982).  Nonetheless, Brickhill  (1991) proposed the convective driving
mechanism  as being responsible  for the  overstability of  g-modes in
DAVs  (see  also Goldreich  \&  Wu  1999).   Although both  mechanisms
predict roughly the observed blue  edge of the instability strip, none
of them  are capable  to yield  the red edge,  where pulsations  of DA
white dwarfs seemingly cease in a very abrupt way (Kanaan 1996).
  
A longstanding problem in the study of pulsating DA white dwarfs is to
find  the reason  of  why only  a  very reduced  number  of modes  are
observed,  as  compared  with  the  richness  of  modes  predicted  by
theoretical  studies. Indeed,  it has  been long  suspected  that some
filtering mechanism must be  acting quite efficiently. The explanation
commonly proposed is that of ``mode trapping'' phenomenon
\footnote{Other  physically plausible mechanism  of mode  filtering is
related to the interaction of convection with pulsation (see Gautschy
et al. 1996).} (Winget et al. 1981; Brassard
et al.  1992a; Bradley 1996). According to this mechanism, those modes
(the trapped  ones) having a  local radial wavelength  comparable with
the thickness of the hydrogen  envelope, require low kinetic energy to
reach  observable  amplitudes.  Then,  most  of  the observed  periods
should  correspond  to  trapped  modes.   Nevertheless,  in  a  recent
asteroseismological  study  of the  ZZ  Ceti  star G117-B15A  (Bradley
1998), the observed period of 215 s, which has the larger amplitude in
the power spectrum, does not  correspond to the trapped mode predicted
by the best fitting model.

The  exploration   of  these  very  important   aspects  requires  the
construction   of  detailed  DA   white  dwarf   evolutionary  models,
particularly  regarding  the   treatment  of  the  chemical  abundance
distribution.   Work  in  this  direction  has recently  begun  to  be
undertaken.  In  fact, Althaus  et al.  (2002)  have carried  out full
evolutionary  calculations  which  take  into account  time  dependent
element diffusion, nuclear burning and  the history of the white dwarf
progenitor in a self-consistent  way. Specifically, these authors have
followed the evolution  of an initially 3 \msun  stellar model all the
way from the stages of hydrogen and helium burning in the core through
the  thermally pulsing  and  mass  loss phases  till  the white  dwarf
state. Althaus  et al. (2002) find  that the shape of  the Ledoux term
(an important ingredient in the computation of the Brunt-V\"ais\"al\"a
frequency;  see Brassard et  al. 1991,  1992ab) is  markedly different
from  that   found  in  previous  detailed  studies   of  white  dwarf
pulsations.  This  is due  partly to the  effect of smoothness  in the
chemical  abundance distribution  caused by  element  diffusion, which
gives rise to less pronounced peaks in the Ledoux term. This, in turn,
leads  to a  substantially  weaker  mode trapping  effect,  as it  has
recently been  found by  C\'orsico et al.  (2001). These  authors have
presented the  first results regarding the trapping  properties of the
Althaus et al. (2002) evolutionary models.

The present work  is designed to explore at  some length the pulsation
properties of  the Althaus et  al.  (2002) evolutionary models  and to
compare  our  predictions  with  those of  others  investigators.   In
addition,  the work  is intended  to bring  some more  insight  to the
phenomenon of  mode trapping  in the frame  of these  new evolutionary
models. In particular, we shall restrict ourselves to analyse the same
stellar     model    as     that    studied     in     C\'orsico    et
al.  (2001). Specifically,  the model,  which belongs  to the  ZZ Ceti
instability  strip, is  analyzed in  the frame  of  linear, non-radial
stellar pulsations  in the  adiabatic approximation. Emphasis  will be
placed  on  assessing  the   role  played  by  the  internal  chemical
stratification in  the behaviour  of eigenmodes, and  the expectations
for the full spectrum of  periods.  Specifically, we shall explore the
effects of the chemical  interfaces on the kinetic energy distribution
of the  modes and their ability  to modify the properties  of the 
g-mode propagation  throughout the star interior.  We  want to mention
that,  because  of the  high  computational  demands  involved in  the
evolutionary calculations in which  white dwarf cooling is assessed in
a self-consistent  way with element  diffusion and the history  of the
pre-white dwarf, we  are forced to restrict our  attention to only one
value for the stellar mass.

The article  is organized as follows.  In  Sect. \ref{sec:compu}, we
briefly  describe our  evolutionary  and pulsational  codes.  In  this
section  we  also discuss  some  aspects  concerning the  evolutionary
properties of  our models.  In  Sect. \ref{sec:resul} we  present in
detail  the pulsation  results. Finally,  Sect.  \ref{sec:conclu} is
devoted to summarizing our findings.

\section{Details of computations} \label{sec:compu}

Our selected ZZ Ceti model  on which the pulsational results are based
has been calculated  by means of an evolutionary  code developed by us
at La Plata  Observatory.  The code, which is based  on a detailed and
up-to-date physical  description, has enabled us to  compute the white
dwarf evolution in a self-consistent  way with the predictions of time
dependent element  diffusion, nuclear burning  and the history  of the
white dwarf  progenitor. The  constitutive physics includes:  new OPAL
radiative opacities for different metallicities, conductive opacities,
neutrino  emission  rates  and  a  detailed  equation  of  state.   In
addition, a  network of 30  thermonuclear reaction rates  for hydrogen
burning (proton-proton chain and  CNO bi-cycle) and helium burning has
been considered.   Nuclear reaction rates  are taken from  Caughlan \&
Fowler (1988) and Angulo et al. (1999) for the $^{12}{\rm C}(\alpha,
\gamma)^{16}{\rm O}$ reaction rate. This  rate is about twice as large
as  that of Caughlan  \& Fowler  (1988).  Abundance  changes resulting
from  nuclear burning  are computed  by means  of a  standard implicit
method of integration.  In particular,  we follow the evolution of the
chemical  species  $^{1}$H,  $^{3}$He, $^{4}$He,  $^{7}$Li,  $^{7}$Be,
$^{12}$C, $^{13}$C,  $^{14}$N, $^{15}$N, $^{16}$O,  $^{17}$O, $^{18}$O
and  $^{19}$F.  Convection  has  been treated  following the  standard
mixing length  theory (B\"ohm-Vitense 1958) with  the mixing-length to
pressure scale  height parameter of $\alpha=  $1.5.  The Schwarzschild
criterium was used to  determine the boundaries of convective regions.
Overshooting  and semi-convection were  not considered.   Finally, the
various processes relevant for  element diffusion have also been taken
into account.  Specifically, we considered the gravitational settling,
and  the chemical and  thermal diffusion  of nuclear  species $^{1}$H,
$^{3}$He,   $^{4}$He,  $^{12}$C,   $^{14}$N  and   $^{16}$O.   Element
diffusion is based on the treatment for multicomponent gases developed
by  Burgers  (1969).  It  is  important to  note  that  by using  this
treatment of diffusion  we are avoiding the widely  used trace element
approximation (see Tassoul et al. 1990). After computing
the  change of  abundances by  effect of  diffusion, they  are evolved
according  to the  requirements  of nuclear  reactions and  convective
mixing.    Radiative  opacities   are  calculated   for  metallicities
consistent with  the diffusion predictions.   This is done  during the
white  dwarf   regime  in   which  gravitational  settling   leads  to
metal-depleted outer layers.  In  particular, the metallicity is taken
as two  times the abundance of  CNO elements.  For  more details about
this and other computational details we refer the reader to Althaus et
al. (2002) and Althaus et al. (2001).

We started the evolutionary calculations from a 3 \msun~ stellar model
at  the zero-age main  sequence.  The  adopted initial  metallicity is
$Z$= 0.02  and the  initial abundance by  mass of hydrogen  and helium
are,   respectively,   $X_{\rm    H}$=   0.705   and   $X_{\rm   He}$=
0.275. Evolution has  been computed at  constant stellar mass  all the
way from the  stages of hydrogen and helium burning in  the core up to
the tip  of the  asymptotic giant branch  where helium  thermal pulses
occur.  After experiencing  11 thermal pulses, the model  is forced to
evolve  towards the  white dwarf  state by  invoking strong  mass loss
episodes.   The adopted  mass loss  rate was  $\approx  10^{-4}$ \msun
yr$^{-1}$  and it  was  applied  to each  stellar  model as  evolution
proceeded.  After the convergence of each new stellar model, the total
stellar mass is  reduced according to the time step  used and the mesh
points are appropriately adjusted.  As a result of mass loss episodes,
a white dwarf  remnant of 0.563 \msun~ is  obtained.  The evolution of
this remnant is pursued through the stage of planetary nebulae nucleus
till  the domain  of the  ZZ  Ceti stars  on the  white dwarf  cooling
branch.

As well known, the shape  of the composition transition zones plays an
important role in  the pulsational properties of DAV  white dwarfs. In
this  sense, an important  aspect of  these calculations  concerns the
evolution of the chemical abundance  during the white dwarf regime. In
particular,  element  diffusion  makes  near  discontinuities  in  the
chemical profile  at the start  of the cooling branch  be considerably
smoothed out by the time the ZZ Ceti domain is reached (see Althaus et
al.   2002).  The  chemical  profile throughout  the  interior of  our
selected white dwarf model is depicted  in the upper panel of Fig.  1.
Only the most  abundant isotopes are shown.  In  particular, the inner
carbon-oxygen core emerges from the convective helium core burning and
from  the   subsequent  stages  in  which   the  helium-burning  shell
propagates outwards.   Note also  the flat profile  of the  carbon and
oxygen  distribution towards  the centre.   This  is a  result of  the
chemical rehomogenization  of the  innermost zone of  the star  due to
Rayleigh-Taylor  instability (see  Althaus  et al.   2002). Above  the
carbon-oxygen interior there is a shell rich in both carbon ($\approx$
35\%) and  helium ($\approx$ 60\%), and an  overlying layer consisting
of nearly pure helium of mass  0.003 \msun.  The presence of carbon in
the helium-rich region below the pure  helium layer is a result of the
short-lived convective  mixing which  has driven the  carbon-rich zone
upwards during  the peak  of the last  helium pulse on  the asymptotic
giant branch.  We want to mention that the total helium content within
the star once helium  shell burning is eventually extinguished amounts
to 0.014  \msun~ and  that the mass  of hydrogen  that is left  at the
start  of the cooling  branch is  about $  1.5 \times  10^{-4}$ \msun,
which is reduced to $7 \times  10^{-5}$ \msun~ due to the interplay of
residual nuclear burning and element diffusion by the time the ZZ Ceti
domain  is reached.  Finally,  we note  that  the inner  carbon-oxygen
profile of  our models is somewhat  different from that  of Salaris et
al. (1997).  In particular, we find that the size of the carbon-oxygen
core is  smaller than  that found  by Salaris et  al. (1997)  with the
consequent result that the drop in the oxygen abundance above the core
is not so pronounced as in the case found by these authors. We suspect
that  this  different behaviour  could  be  a  result of  a  different
treatment of the convective boundary during the core helium burning.

For  the pulsation  analysis we  have employed  the code  described in
C\'orsico \& Benvenuto (2002).  We  refer the reader to that paper for
details.  Here  we shall describe briefly our  strategy of calculation
and  mention the pulsation  quantities computed  that are  relevant in
this   study.   The  pulsational   code  is   based  on   the  general
Newton-Raphson technique  (like the Henyey method  employed in stellar
evolution  studies).   The  code  solves  the  differential  equations
governing the  linear, non-radial stellar pulsations  in the adiabatic
approximation (see Unno et al.  1989 for details of their derivation).
The boundary  conditions at the  stellar centre and surface  are those
given by Osaki  \& Hansen (1973) (see Unno et  al.  1989 for details).
Following   previous   studies   of   white  dwarf   pulsations,   the
normalization condition  adopted is $\delta r  / r= 1$  at the stellar
surface.   After  selecting  a  starting  stellar model  we  choose  a
convenient  period window  and  the interval  of  interest in  $T_{\rm
eff}$. The  evolutionary code computes  the white dwarf  cooling until
the  hot edge  of the  $T_{\rm eff}$-interval  is reached.   Then, the
program  calls the set  of pulsation  routines to  begin the  scan for
modes.    In  order  to   obtain  the   first  approximation   to  the
eigenfunctions and the eigenvalue of a mode we have applied the method
of the  discriminant (Unno et  al.  1989). Specifically, we  adopt the
potential  boundary condition  (at  the surface)  as the  discriminant
function (see C\'orsico \& Benvenuto 2002).  When a mode is found, the
code generates  an approximate solution which  is iteratively improved
to   convergence   (of   the   eigenvalue   and   the   eigenfunctions
simultaneously) and then stored.  This procedure is repeated until the
period interval is covered.  Then, the evolutionary code generates the
next  stellar model  and  calls pulsation  routines  again.  Now,  the
previously  stored modes  are taken  as initial  approximation  to the
modes of the present stellar model and iterated to convergence.

For each computed mode we obtain the eigenperiod $P_k$ ($P_k= 2 \pi /
\sigma_k  $, being  $\sigma_k$ the  eigenfrequency)  and the dimensionless
eigenfunctions  $y_1,\cdots, y_4$  (see Unno  et al.   1989  for their
definition).    With  these   eigenfunctions  and   the  dimensionless
eigenvalue $\omega_k^2  = \sigma^2_k (G M_* /  R_*^3)^{-1}$ we compute
for  each mode  considered  the oscillation  kinetic energy,  $(E_{\rm
kin})_k$, given by:

\begin{equation} \label{eq1}
(E_{\rm kin})_k = \frac{1}{2} (G M_* R_*^2) \omega_k^2 \int_{0}^{1} 
x^2 \rho \left[ x^2 y_1^2 + x^2 \frac{\ell (\ell +1)}{(C_1 \omega_k^2)^2}
y_2^2\right] dx,
\end{equation}

\noindent   and   the  first   order   rotation  splitting   coefficient,
$C_{\ell,k}$, 

\begin{equation} \label{eq2}
C_{\ell,k}= \frac{(G M_* R_*^2)}{2 (E_{\rm kin})_k}
\int_{0}^{1} 
\frac{x^2 \rho}{C_1} \left[ 2 x^2 y_1 y_2 + \frac{x^2}{C_1 \omega_k^2}
y_2^2\right] dx,
\end{equation}

\noindent where $M_*$  and $R_*$ are the stellar  mass and the stellar
radius respectively, $G$ is  the gravitation constant, $C_1= (r/R_*)^3
(M_*/M_r)$  and $x  = r  / R_*$.  In addition,  we compute  the weight
functions, $WF$,  and the variational period, $P_k^{\rm  V}$, as given
by Kawaler et al. (1985).  Finally, for each model computed
we derive the asymptotic spacing of periods, $\Delta P_{\rm A}$, given
by (Tassoul 1980; Tassoul et al. 1990):

\begin{equation} \label{eq3}
\Delta P_{\rm A} = \frac{P_0}{\sqrt{\ell(\ell+1)}},
\end{equation}

\noindent and $P_0$ is defined as

\begin{equation} \label{eq4}
P_0 = 2 \pi^2\ \left[\int_{0}^{x_2} \frac{N}{x} dx \right]^{-1},
\end{equation}

\noindent   where $x_2$ correspond to the location of the base of 
the outer convection zone.  The Brunt-V\"ais\"al\"a frequency ($N$), a
fundamental quantity of white  dwarf pulsations, is computed employing
the ``modified Ledoux'' treatment.  This treatment explicitly accounts
for the  contribution to $N^2$ from  any change in  composition in the
interior of model  (the zones of chemical transition)  by means of the
Ledoux term $B$ (see Brassard et  al.  1991).  We want to mention that
we have also employed a numerical differentiation scheme for computing
$N^2$ directly from its definition.   We found that this scheme yields
the same results as those derived from the modified Ledoux treatment.

As  mentioned,  for the  pulsation  analysis  in  this study  we  have
selected a white dwarf model representative of the ZZ Ceti instability
band.   Specifically, we  have picked  out a  0.563  ${\rm M_{\odot}}$
model at  $T_{\rm eff}\approx$ 12000  K. In Table  1 we show  the main
characteristics  of   our  template   model.   In  the   interests  of
comparison, the same quantities corresponding to a similar model of P.
Bradley    (2002)   (private    communication)    are   shown.     The
Brunt-V\"ais\"al\"a frequency and  the Ledoux term ($B$) corresponding
to our  template model are  shown in the  middle and bottom  panels of
Fig.   1 in terms  of the  outer mass  fraction.  Note  the particular
shape of $B$, which is a
\begin{center}
\centerline{Table 1. {}}
\begin{tabular}{lcc}
\hline
\hline
                &Our template & Bradley's model\\
                & model & \\
\hline
$M_*/{\rm M_{\odot}}$        &    0.563   &  0.560   \\
$T_{\rm eff}$ [K]            &    11996   &  12050   \\ 
$\log(L_*$/L$_{\odot})$      &    -2.458  &  -2.462  \\ 
$\log(R_*$/R$_{\odot})$      &    -1.864  &  -1.866  \\
$\log(M_{\rm H}/M_*)$        &    -3.905  &  -3.824  \\
$\log(M_{\rm He}/M_*)$       &    -1.604  &  -1.824  \\
$\log(\rho_c)$ [g cm$^{-3}$] &    6.469   &  6.466   \\
$\log(T_c)$ [10$^6$ K]       &    7.086   &  7.087   \\
\hline
\end{tabular}
\end{center}
direct consequence of the chemical  profile.  In turn, the features of
$B$ are reflected in the  Brunt-V\" ais\"al\"a frequency.  As a result
of element diffusion, the chemical profiles of our evolutionary models
are very  smooth in the interfaces  (see upper panel of  Fig.  1), and
this  explains the presence  of extended  tails in  the shape  of $B$.
Also, note that our model  is characterized by a chemical interface in
which three ionic species  in appreciable abundances coexists: helium,
carbon and oxygen (see early in this section).  This transition region
gives  two contributions  to  $B$,  one of  them  of relatively  great
magnitude, placed at $\log q  \sim -1.4$, and the other, more external
and of very low  height at $\log q \sim -2.2$.  As  a last remark, the
contribution of the hydrogen-helium interface to $B$ is less than that
corresponding to  the helium-carbon-oxygen transition.   Note that the
contributions from the Ledoux term are translated into smooth bumps on
$N^2$.   The characteristics  of $B$  and  $N^2$ as  predicted by  our
models are markedly different from  those found in previous studies in
which  the white  dwarf  evolution  is treated  in  a simplified  way,
particularly  regarding  the chemical  abundance  distribution in  the
outer  layers (e.g.,  Tassoul et  al.   1990; Brassard  et al.   1991,
1992ab; Bradley 1996).  For more details see Althaus et al. (2002).

\section{Results} \label{sec:resul}

For our template model we have  computed g-modes with $\ell= 1, 2$ and
$3$  (we   do  this   because  geometric  cancellation   effects  grow
progressively  for  larger  $\ell$  in  non-radial  oscillations;  see
Dziembowski 1977),  with periods  in the range  of 50 s  $\lesssim P_k
\lesssim$  1300 s. Let  us quote  that for  mode calculations  we have
employed  up  to   5000  mesh  points.   For  all   of  our  pulsation
calculations, the relative difference  between $P_k$ and $P_k^{\rm V}$
remains lower than $10^{-3}$. This gives an indication of the accuracy
of our calculations.

We begin by examining Figs. 2 to 4, the upper panels of which show the
logarithm   of  the   oscillation  kinetic   energy  of   modes  with,
respectively,  $\ell= 1,  2$ and  $3$  in terms  of computed  periods.
Middle panels depict the values for the forward period spacing $\Delta
P_k$  ($\equiv P_{k+1}  -  P_k$) together  with  the asymptotic  value
$\Delta P_{\rm A}$ as given by dotted lines
\footnote  {Strictly   speaking,  $\Delta  P_{\rm  A}$   as  given  by
Eq. (\ref{eq3}) corresponds to the asymptotic (high $k$) separation of
periods for  {\it chemically  homogeneous} and {\it  radiative} stars.
However, it  is very close to  the {\it mean period  spacing} even for
chemically stratified  white dwarfs.  See  Tassoul et al.   (1990) for
details.}.  Finally,  in the bottom  panel of these figures  we depict
the  $C_{\ell,k}$ values  as  well as  the  asymptotic values  (dotted
lines)  that these  coefficients  adopt for  high  overtones, that  is
$C_{\ell,k}  \approx  1 /  \ell  (\ell  +  1)$ (Brickhill  1975).   An
inspection  of  plots reveals  some  interesting characteristics.   To
begin  with,  the quantities  plotted  exhibit  two clearly  different
trends.  Indeed, for $P_k \gtrsim 500 - 600$ s and irrespective of the
value  of $\ell$, the  distribution of  oscillation kinetic  energy is
quite smooth.  Note that  the $\log(E_{\rm kin})_k$ values of adjacent
modes are quite similar, which is in contrast with the situation found
for lower  periods.  On  the other hand,  the period  spacing diagrams
show  appreciable  departures  of  $\Delta P_k$  from  the  asymptotic
prediction (Eq.   3) for $P_k \lesssim  500 - 600$ s.   As well known,
this is due  mostly to the presence of  chemical abundance transitions
in DA white dwarfs.  In  contrast, for higher periods the $\Delta P_k$
of  the  modes  tend to  $\Delta  P_{\rm  A}$.   Also, note  that  the
$C_{\ell,k}$ values tend to the  asymptotic value for $P_k \gtrsim 500
- 600$ s.

An  important aspect  of  the present  study  is related  to the  mode
trapping  and confinement properties  of our  models. For  the present
analysis we  shall employ the  weight functions, $WF$.  We  elect $WF$
because this function gives the relative contribution of the different
regions  in the star  to the  period formation  (Kawaler et  al. 1985;
Brassard  et  al. 1992ab).   We  want to  mention  that  we have  also
carefully  examined the density  of kinetic  energy (the  integrand of
Eq. 1)  for each computed mode.  For our purposes  here, this quantity
gives us basically the same information that provided by $WF$. We show
in Fig.  5  to 7 the $WF$ for all of  the computed modes corresponding
to $\ell=1$.   In addition, we include  in each plot  of these figures
the Ledoux  term in  arbitrary units (dotted  lines) in order  to make
easier the location  of the chemical transition regions  of the model.
In the interests of a proper interpretation of these figures, we suggest
the  reader  to see  also  Fig.   2. For  low  periods,  a variety  of
behaviour   is  encountered.    For  instance,   the  g$_1$   mode  is
characterized by a $WF$ corresponding  to the well known mode trapping
phenomenon,  that  is,  g$_1$  is  formed in  the  very  outer  layers
irrespective  of  the  details  of  the deeper  chemical  profile,  as
previously reported by previous studies (see Brassard et al.  1992ab).
In  contrast, it  is the  helium-carbon-oxygen transition  that mostly
contributes  to   the  formation  of   the  g$_2$  mode,   whilst  the
hydrogen-helium  transition plays a  minor role.   This mode  would be
representative of the ``confined  modes'' according to Brassard et al.
(1992ab).  $WF$ for modes g$_3$  and g$_4$ is qualitatively similar to
that  of g$_1$, except  that they  are not  exclusively formed  in the
hydrogen-rich   envelope,  but   also   in  the   helium-carbon-oxygen
interface.  On  the other hand,  the high-density zone  underlying the
helium-carbon-oxygen transition plays a major role in the formation of
mode g$_5$.  From Eq.  (1)  is clear that the $(E_{\rm kin})_k$ values
are  proportional  to  the  integral of  the  squared  eigenfunctions,
weighted by $\rho$. As a result,  the g$_5$ mode is characterized by a
high oscillation  kinetic energy value  (see Fig.  2).  Note  that the
helium-carbon-oxygen  transition   region  also  contributes   to  the
formation  of  modes  g$_6$   and  g$_{10}$.   The  g$_{10}$  mode  is
particularly interesting,  because it is  formed over a wide  range of
the stellar interior, thus being also a high kinetic energy mode.  The
$WF$s  corresponding  to remaining  modes  do  not differ  appreciably
amongst them.  They exhibit contributions mainly from the outer layers
of  the  model, though  they  also  show  small amplitudes  in  deeper
regions.  Note that for  all of the modes shown in Figs.  5 to 7 there
is a strong contribution  to $WF$s from the hydrogen-helium transition
region.   This  indicate that,  as  found  in  previous studies,  this
chemical interface plays a fundamental role in the period formation of
modes.  We want to mention that  we have elected for this analysis the
dipolar modes  ($\ell=1 $) for brevity;  the results for  $\ell= 2, 3$
are qualitatively similar to those of $\ell= 1$.

From the  analysis performed above  based on the weight  functions, we
can clearly  appreciate that for $P_k  \gtrsim 500 - 600$  s the outer
layer  of the  model  appreciably  contribute to  the  $WF$s. This  is
expected,  because,  as well  known,  $g$-modes  in  white dwarfs  are
envelope modes. As  mentioned, the $WF$s of high  order modes are very
similar,  indicating  that  these  modes  have  essentially  the  same
characteristics.   At this  point,  we could,  in principle,  classify
these  modes either  like trapped  or partially  trapped in  the outer
envelope or like  ``normal'' modes (in the terminology  of Brassard et
al.   1992),  that  is,  without enhanced  or  diminished  oscillation
kinetic  energies  as  in  the  case of  eigenmodes  corresponding  to
chemically   homogeneous  stellar   models.   In   fact,   the  curves
$\log(E_{\rm kin})_k -  P_k$ depicted in Figs.  2  to 4, in particular
for periods  exceeding $500  - 600$ s,  strongly resemble  the kinetic
energy distribution corresponding  to a model in which  there no exist
chemical interfaces.   With the aim  of solving such an  ambiguity, we
have performed  pulsation calculations  arbitrarily setting $B=  0$ in
the computation  of the Brunt-V\"ais\"al\"a  frequency.  As mentioned,
the  modified Ledoux  treatment  employed in  the  computation of  the
Brunt-V\"ais\"al\"a frequency bears explicitly the effect from changes
in  chemical composition  by means  of the  Ledoux term  $B$.   So, by
forcing $B=  0$ the effects  of the chemical transitions  are strongly
minimized (but not completely eliminated; see inset of Fig. 1c) on the
whole  pulsational  pattern. In  this  way  we  obtain an  approximate
chemically homogeneous  white dwarf model (see Brassard  et al.  1992b
for a  similar numerical  experiment). The oscillation  kinetic energy
values resulting for this simulated ``homogeneous'' model are shown in
Fig.  8 with  dotted lines. In the interests of  a comparison, we show
the  results corresponding  to our  (full) template  model  with solid
lines.  It is clear from  the figure that the distribution of $(E_{\rm
kin})_k$ values for  both sets of computations (and  for each value of
$\ell$) is very  similar in the region of  long periods. However, note
that the  curves corresponding  to the modified  model are  shifted to
higher energies (by $ \approx 0.2$ dex) as compared with the situation
of the full model.  We have  carefully compared the $WF$ for each mode
of the full  model with the corresponding mode  of the ``homogeneous''
model (i.e. modes which  have closest period values although generally
for different radial order $k$). We found that, for modes with periods
exceeding  $\approx 600$  s, the  $WF$s are  almost identical  in both
cases at  the regions above the  hydrogen-helium transition.  However,
below this  interface the  $WF$s corresponding to  the ``homogeneous''
model show  larger amplitudes  as compared with  the case of  the full
model. Thus,  we can conclude that  for the full model,  {\it all} the
modes  corresponding to  the  long period  region  of the  pulsational
spectrum must be considered  as partially trapped in the hydrogen-rich
envelope.   In   others  words,  the  chemical   distribution  at  the
hydrogen-helium  transition has  noticeable effect  on each  mode, but
this effect is the same  for all modes.  This conclusion is reinforced
by  the fact  that  the first  order  rotation splitting  coefficients
($C_{\ell,k}$) for the full model adopt higher values as compared with
those corresponding  to the ``homogeneous'' model (not  shown here for
brevity), thus  lying nearest to the asymptotic  prediction.  As found
by  Brassard et  al.   (1992ab), it  is  an additional  characteristic
feature of  trapped modes in  the hydrogen-rich outer region  of white
dwarfs.

An important finding of this  work is the effect of chemical abundance
distribution  resulting  from time  dependent  diffusion  on the  mode
trapping properties in  DA white dwarfs.  In fact, as  shown in Fig. 2
to 4, for periods exceeding $\approx 500 - 600$ s, the distribution of
$(E_{\rm  kin})_k$ is  smooth, and  $\Delta  P_k$ values  tend to  the
asymptotic value.  This is quite different from that found in previous
studies.   Our  calculations  reveal   that  the  capability  of  mode
filtering  due to  mode  trapping effects  virtually  vanish for  high
periods when  account is made  of white dwarf models  with diffusively
evolving chemical  stratifications (see  C\'orsico et al.   2001).  In
order to make  a detailed comparison of the  predictions of our models
with those  found in previous  studies we have carried  out additional
pulsational  calculations  by assuming  diffusive  equilibrium in  the
trace  element  approximation at  the  hydrogen-helium interface  (see
Tassoul et  al.  1990).  This  treatment has been commonly  invoked in
most  of the  pulsation studies  to model  the  composition transition
regions.    The   resulting   hydrogen   chemical  profile   and   the
corresponding  Ledoux term and  Brunt-V\"ais\"al\"a frequency  $N$ are
shown  in Fig.   9, together  with the  predictions of  time dependent
element diffusion.   The trace element  assumption leads to  an abrupt
change in the  slope of the chemical profile  which is responsible for
the pronounced peak in  the Brunt-V\"ais\"al\"a frequency at $\log(1 -
M_r / M_*) \approx -4$.  As can be clearly seen in Fig.  10 for $\ell=
1$ to 3, the diffusive  equilibrium in the trace element approximation
gives  rise  to an  oscillation  kinetic  energy  spectrum and  period
spacing distribution that are substantially different from those given
by the full  treatment of diffusion (see Figs.   2 to 4), particularly
for high periods. The most  outstanding feature depicted by Fig. 10 is
the  trapping  signatures  exhibited  by  certain modes  both  in  the
$\log(E_{\rm kin})_k$  and $\Delta P_k$ values.  This  is in agreement
with other  previous results (see Brassard et  al. 1992b, particularly
their  figures 20a  and  21a for  the  case of  $M_{\rm  H} =  10^{-4}
M_*$)\footnote {Note that,  as found by Brassard et  al.  (1992a), the
contrast between the kinetic energies  values of the trapped modes and
those of the  non-trapped ones is not very  large for massive hydrogen
envelopes (like that presented in this work), due to the fact that the
hydrogen-helium  interface is  located  in a  deep, highly  degenerate
region, where  the amplitudes of  eigenfunctions are very  small.}. As
well  known,  trapped  modes  correspond  to  those  modes  which  are
characterized by minima in their oscillation kinetic energy values and
local  minima in  the  period  spacing having  the  same $k$-value  or
differing by 1.   For the purpose of illustration,  we compare in Fig.
11  and 12  the  predictions  of equilibrium  diffusion  in the  trace
element   approximation    and   time-dependent   element   diffusion,
respectively,  for  $WF$  corresponding  to  the  modes  g$_{38}$  and
g$_{39}$  with   $\ell=  2$.   Clearly,  in  the   case  of  diffusive
equilibrium  in   the  trace  element   approximation,  mode  g$_{39}$
corresponds to  a trapped  mode characterized by  small values  of the
weight function below the hydrogen-helium transition, as compared with
the adjacent, non-trapped mode g$_{38}$.  By contrast, such modes show
very similar amplitudes of their $WF$  when account is taken of a full
diffusion treatment  to model the composition  transition regions (see
Fig.  12).   We would  also  like  to comment  on  the  fact that  the
diffusive  equilibrium condition  is  far from  being  reached at  the
bottom  of the  hydrogen  envelope of  our  model. In  Althaus et  al.
(2002) we argued  that the situation  of diffusive equilibrium  in the
deep  layers of  a  DAV white  dwarf  is not  an  appropriate one  for
describing   the   shape   of   the  chemical   composition   at   the
hydrogen-helium  transition zone. In  fact, during  the ZZ  Ceti stage
time-dependent  diffusion  modifies the  spatial  distribution of  the
elements, particularly  at the chemical  interfaces (see also  Iben \&
MacDonald  1985).   In  addition,  for  the  case  of  thick  hydrogen
envelopes,  we  have  recently  found  that under  the  assumption  of
diffusive equilibrium, a white dwarf does not evolve along the cooling
branch, but rather it experiences a hydrogen thermonuclear shell flash
(see  C\'orsico et al.  2002).  This  is so  because if  diffusion had
plenty of time to evolve to  an equilibrium situation then the tail of
the hydrogen  distribution would  have been able  to reach  hot enough
layers to be ignited in a flash fashion.

To place  some of  the results  of the foregoing  paragraph in  a more
quantitative basis,  we list in Tables  2 and 3 the  values for $P_k$,
$\Delta  P_k$ and  $\log(E_{\rm  kin})_k$ for  modes corresponding  to
$\ell=$ 1 and $\ell=$ 2, in the case of equilibrium diffusion and time
dependent  element diffusion.  In  Table 2,  the ``m''  corresponds to
minima, and  ``M'' stands for maxima.   We have labeled  the minima of
$\Delta P_k$  and the minima  and maxima of $\log(E_{\rm  kin})_k$, in
correspondence with Fig.   10.  Note that for the  case with $\ell= 1$
there is a direct correlation  (indicated by arrows) between minima in
$\log(E_{\rm  kin})_k$ and  $\Delta P_{k-1}$  for most  of  high order
modes,  whereas for  the case  with $\ell=$  2 this  correspondence is
between minima  in $\log(E_{\rm kin})_k$ and $\Delta  P_k$.  The modes
with minima in  kinetic energy are classified as  trapped (T) ones. In
contrast   to  the   case  of   equilibrium  diffusion,   the  results
corresponding to the time dependent element diffusion treatment do not
show clear minima  or maxima in kinetic energy,  as can be appreciated
in Figs. 2 to 4 and Table 3. We have compared the periods of our model
with  those kindly  provided  by Bradley  corresponding  to his  0.560
$M_{\odot}$  white dwarf  model,  and  we find  that  our periods  are
typically  6  \% shorter.  In  part, this  difference  is  due to  the
somewhat smaller mass  of the Bradley's model and  the different input
physics characterizing both stellar models.

\begin{center}
\begin{table*}
\centerline{Table 2.  Pulsation properties for $\ell=  1,  2$ modes}
\centerline{corresponding to the case  of equilibrium diffusion.}
\begin{tabular*}{140mm}{crclcl|crclcl}
\hline
\hline
\multicolumn{6}{c|}{$\ell= 1$} & \multicolumn{6}{c}{$\ell= 2$}\\
\hline
\multicolumn{6}{c|}{$\Delta P_{\rm A}=$ 46.26 s} & \multicolumn{6}{c}{$\Delta P_{\rm A}=$ 26.71 s}\\
\hline
$k$ & $P_k$ && $\Delta P_k$ && $\log(E_{\rm kin})_k$  &
$k$ & $P_k$ && $\Delta P_k$ && $\log(E_{\rm kin})_k$  \\
 & [s] & & [s] && [erg] &
 & [s] & & [s] && [erg] \\
\hline
 1 & 126.99 &T& 76.38   && 45.84 m & 1 & 73.39  &T& 48.02   && 45.84 m \\  
 2 & 203.37 & & 73.91   && 46.97 M & 2 & 121.40 & & 39.77   && 46.85 M \\  
 3 & 277.28 &T& 26.02 m &$\leftarrow$& 44.60 m & 3 & 161.17 &T& 18.11 m &$\leftarrow$& 44.52 m \\  
 4 & 303.30 & & 40.79   && 44.61 m & 4 & 179.28 & & 39.51   && 44.61 M \\  
 5 & 344.09 & & 40.65   && 44.93 M & 5 & 218.80 &T& 14.17 m &$\leftarrow$& 43.77 m \\  
 6 & 384.75 & & 44.00   && 43.62 m & 6 & 232.97 & & 17.06   && 43.89 M \\  
 7 & 428.75 & & 61.25   && 43.43   & 7 & 250.03 & & 33.50   && 43.49   \\  
 8 & 490.00 &T& 59.93   && 42.90 m & 8 & 283.53 & & 40.06   && 42.89   \\  
 9 & 549.93 & & 21.12 m &$\swarrow$& 42.99 M & 9 & 323.59 &T& 24.67   && 42.53 m \\  
10 & 571.05 & & 47.22   && 42.59   & 10 & 348.26 & & 18.24 m &$\swarrow$& 42.48 M \\  
11 & 618.27 & & 44.54 m && 42.23   & 11 & 366.50 & & 19.88   && 42.27   \\  
12 & 662.81 &T& 48.65   &$\nwarrow$& 41.91 m & 12 & 386.39 & & 35.55   && 41.95   \\  
13 & 711.46 & & 30.69 m && 42.24 M & 13 & 421.94 &T& 21.12 m &$\leftarrow$& 41.76 m \\  
14 & 742.15 &T& 46.66   &$\nwarrow$& 41.75 m & 14 & 443.06 & & 30.37   && 41.96 M \\  
15 & 788.81 & & 42.28 m && 41.98 M & 15 & 473.43 &T& 19.88 m &$\leftarrow$& 41.58 m \\  
16 & 831.09 &T& 48.51   &$\nwarrow$& 41.48 m & 16 & 493.32 & & 25.75   && 41.66 M \\  
17 & 879.60 & & 43.99 m && 41.55 M & 17 & 519.06 & & 19.59 m && 41.45   \\  
18 & 923.59 &T& 48.15   &$\nwarrow$& 41.16 m & 18 & 538.65 & & 30.84   && 41.21   \\  
19 & 971.75 & & 33.31 m && 41.31 M & 19 & 569.49 &T& 22.68 m &$\leftarrow$& 41.10 m \\  
20 & 1005.05 &T& 47.02  &$\nwarrow$& 41.16 m & 20 & 592.18 & & 27.64   && 41.29 M \\  
21 & 1052.07 & & 42.38 m && 41.32 M & 21 & 619.82 & & 19.68 m && 41.22  \\  
22 & 1094.45 &T& 55.95   &$\nwarrow$& 41.10 m & 22 & 639.51 & & 32.53   && 41.26 M \\  
23 & 1150.40 & & 32.44 m && 41.22 M & 23 & 672.04 &T& 24.88 m &$\leftarrow$& 41.00 m \\  
24 & 1182.83 &T& 43.15   &$\nwarrow$& 41.09 m & 24 & 696.92 & & 27.97   && 41.12 M \\  
25 & 1225.98 & & 40.36 m && 41.16 M & 25 & 724.89 &T& 21.88 m &$\leftarrow$& 40.93 m \\  
26 & 1266.34 &T& 53.93   &$\nwarrow$& 40.93 m & 26 & 746.77 & & 25.83   && 41.25 M \\  
27 & 1320.27 & & 38.87 m && 41.11 M & 27 & 772.60 &T& 23.68 m &$\leftarrow$& 41.01 m \\  
$\ldots$&$\ldots$&&$\ldots$&&$\ldots$ & 28 & 796.28 & & 28.33   && 41.18 M \\  
$\ldots$&$\ldots$&&$\ldots$&&$\ldots$ & 29 & 824.61 &T& 22.94 m &$\leftarrow$& 41.03 m \\  
$\ldots$&$\ldots$&&$\ldots$&&$\ldots$ & 30 & 847.55 & & 27.53   && 41.17 M \\  
$\ldots$&$\ldots$&&$\ldots$&&$\ldots$ & 31 & 875.08 &T& 21.89 m &$\leftarrow$& 41.04 m \\  
$\ldots$&$\ldots$&&$\ldots$&&$\ldots$ & 32 & 896.97 & & 30.30   && 41.14 M \\  
$\ldots$&$\ldots$&&$\ldots$&&$\ldots$ & 33 & 927.27 &T& 25.47 m &$\leftarrow$& 40.98 m \\  
$\ldots$&$\ldots$&&$\ldots$&&$\ldots$ & 34 & 952.74 & & 27.78   && 41.22 M \\  
$\ldots$&$\ldots$&&$\ldots$&&$\ldots$ & 35 & 980.52 &T& 23.58 m &$\leftarrow$& 41.06 m \\  
$\ldots$&$\ldots$&&$\ldots$&&$\ldots$ & 36 & 1004.10 & & 26.44  && 41.32 M \\  
$\ldots$&$\ldots$&&$\ldots$&&$\ldots$ & 37 & 1030.54 &T& 23.93 m &$\leftarrow$& 41.12 m \\  
$\ldots$&$\ldots$&&$\ldots$&&$\ldots$ & 38 & 1054.46 & & 28.40   && 41.35 M \\  
$\ldots$&$\ldots$&&$\ldots$&&$\ldots$ & 39 & 1082.87 &T& 23.20 m &$\leftarrow$& 41.18 m \\  
$\ldots$&$\ldots$&&$\ldots$&&$\ldots$ & 40 & 1106.07 & & 26.52   && 41.48 M \\  
$\ldots$&$\ldots$&&$\ldots$&&$\ldots$ & 41 & 1132.58 &T& 23.99 m &$\leftarrow$& 41.27 m \\  
$\ldots$&$\ldots$&&$\ldots$&&$\ldots$ & 42 & 1156.57 & & 29.72   && 41.46 M \\  
$\ldots$&$\ldots$&&$\ldots$&&$\ldots$ & 43 & 1186.29 &T& 25.32 m &$\leftarrow$& 41.32 m \\  
$\ldots$&$\ldots$&&$\ldots$&&$\ldots$ & 44 & 1211.61 & & 27.85   && 41.57 M \\  
$\ldots$&$\ldots$&&$\ldots$&&$\ldots$ & 45 & 1239.47 &T& 24.74 m &$\leftarrow$& 41.41 m \\  
$\ldots$&$\ldots$&&$\ldots$&&$\ldots$ & 46 & 1264.20 & & 26.62   && 41.68 M \\  
$\ldots$&$\ldots$&&$\ldots$&&$\ldots$ & 47 & 1290.82 &T& 25.45 m &$\leftarrow$& 41.48 m \\  
$\ldots$&$\ldots$&&$\ldots$&&$\ldots$ & 48 & 1316.27 & & 26.83   && 41.70 M \\  
\hline
\end{tabular*}
\end{table*}
\end{center}

\begin{center}
\begin{table*}
\centerline{Table  3.  Same as Table  2, but  for  the  case of  
time-dependent}
\centerline{element diffusion.}
\begin{tabular*}{110mm}{crcc|crcc}
\hline
\hline
\multicolumn{4}{c|}{$\ell= 1$} & \multicolumn{4}{c}{$\ell= 2$}\\
\hline
\multicolumn{4}{c|}{$\Delta P_{\rm A}=$ 45.83 s} & \multicolumn{4}{c}{$\Delta P_{\rm A}=$ 26.30 s}\\
\hline 
$k$ & $P_k$ & $\Delta P_k$  & $\log(E_{\rm kin})_k$  &
$k$ & $P_k$ & $\Delta P_k$  & $\log(E_{\rm kin})_k$  \\
 & [s] & [s] & [erg] &
 & [s] & [s] & [erg] \\
\hline
 1 & 126.98  & 76.87 & 45.84 &  1 & 72.14  & 49.34 & 45.84 \\ 
 2 & 203.85  & 66.57 & 46.83 &  2 & 121.48 & 34.89 & 46.69 \\ 
 3 & 270.43  & 33.77 & 44.77 &  3 & 156.38 & 21.63 & 44.72 \\ 
 4 & 304.19  & 35.98 & 44.48 &  4 & 178.00 & 32.08 & 44.41 \\ 
 5 & 340.17  & 36.30 & 44.69 &  5 & 210.09 & 23.30 & 44.10 \\ 
 6 & 376.47  & 44.67 & 43.97 &  6 & 233.39 & 10.74 & 43.84 \\ 
 7 & 421.13  & 67.75 & 43.30 &  7 & 244.12 & 36.72 & 43.40 \\ 
 8 & 488.88  & 51.08 & 43.04 &  8 & 280.85 & 31.47 & 43.05 \\ 
 9 & 539.95  & 26.84 & 42.78 &  9 & 312.31 & 33.28 & 42.67 \\ 
10 & 566.80  & 46.27 & 43.04 & 10 & 345.59 & 16.62 & 42.44 \\ 
11 & 613.06  & 44.76 & 42.18 & 11 & 362.21 & 19.97 & 42.30 \\ 
12 & 657.83  & 46.99 & 42.03 & 12 & 382.18 & 30.40 & 42.05 \\ 
13 & 704.82  & 32.45 & 42.08 & 13 & 412.58 & 25.26 & 41.87 \\ 
14 & 737.27  & 42.44 & 41.93 & 14 & 437.84 & 26.30 & 41.86 \\ 
15 & 779.71  & 44.78 & 41.88 & 15 & 464.14 & 24.63 & 41.78 \\ 
16 & 824.49  & 43.42 & 41.68 & 16 & 488.77 & 21.89 & 41.58 \\ 
17 & 867.91  & 50.43 & 41.50 & 17 & 510.66 & 22.90 & 41.59 \\ 
18 & 918.34  & 43.07 & 41.23 & 18 & 533.56 & 24.96 & 41.30 \\ 
19 & 961.42  & 38.91 & 41.20 & 19 & 558.53 & 27.11 & 41.14 \\ 
20 & 1000.33 & 40.24 & 41.28 & 20 & 585.64 & 24.25 & 41.17 \\ 
21 & 1040.57 & 45.71 & 41.19 & 21 & 609.89 & 22.92 & 41.21 \\ 
22 & 1086.28 & 50.32 & 41.19 & 22 & 632.80 & 26.05 & 41.27 \\ 
23 & 1136.60 & 40.88 & 41.13 & 23 & 658.86 & 28.13 & 41.09 \\ 
24 & 1177.48 & 36.91 & 41.17 & 24 & 686.98 & 27.00 & 41.00 \\ 
25 & 1214.40 & 42.99 & 41.06 & 25 & 713.98 & 23.95 & 40.97 \\ 
26 & 1257.38 & 46.02 & 40.96 & 26 & 737.94 & 22.71 & 41.01 \\ 
27 & 1303.40 & 46.39 & 40.99 & 27 & 760.64 & 24.69 & 41.07 \\ 
$\ldots$& $\ldots$& $\ldots$& $\ldots$& 28 & 785.33 & 25.47 & 41.02 \\ 
$\ldots$& $\ldots$& $\ldots$& $\ldots$& 29 & 810.80 & 25.67 & 41.05 \\ 
$\ldots$& $\ldots$& $\ldots$& $\ldots$& 30 & 836.47 & 24.62 & 41.04 \\ 
$\ldots$& $\ldots$& $\ldots$& $\ldots$& 31 & 861.09 & 24.42 & 41.04 \\ 
$\ldots$& $\ldots$& $\ldots$& $\ldots$& 32 & 885.51 & 25.09 & 41.05 \\ 
$\ldots$& $\ldots$& $\ldots$& $\ldots$& 33 & 910.60 & 27.09 & 41.02 \\ 
$\ldots$& $\ldots$& $\ldots$& $\ldots$& 34 & 937.69 & 27.41 & 41.03 \\ 
$\ldots$& $\ldots$& $\ldots$& $\ldots$& 35 & 965.11 & 25.02 & 41.09 \\ 
$\ldots$& $\ldots$& $\ldots$& $\ldots$& 36 & 990.12 & 24.63 & 41.15 \\ 
$\ldots$& $\ldots$& $\ldots$& $\ldots$& 37 & 1014.75 & 24.66 & 41.18 \\ 
$\ldots$& $\ldots$& $\ldots$& $\ldots$& 38 & 1039.41 & 25.64 & 41.21 \\ 
$\ldots$& $\ldots$& $\ldots$& $\ldots$& 39 & 1065.05 & 26.07 & 41.23 \\ 
$\ldots$& $\ldots$& $\ldots$& $\ldots$& 40 & 1091.12 & 24.20 & 41.29 \\ 
$\ldots$& $\ldots$& $\ldots$& $\ldots$& 41 & 1115.32 & 24.33 & 41.35 \\ 
$\ldots$& $\ldots$& $\ldots$& $\ldots$& 42 & 1139.66 & 26.34 & 41.36 \\ 
$\ldots$& $\ldots$& $\ldots$& $\ldots$& 43 & 1166.00 & 27.19 & 41.37 \\ 
$\ldots$& $\ldots$& $\ldots$& $\ldots$& 44 & 1193.19 & 26.48 & 41.42 \\ 
$\ldots$& $\ldots$& $\ldots$& $\ldots$& 45 & 1219.67 & 26.20 & 41.46 \\ 
$\ldots$& $\ldots$& $\ldots$& $\ldots$& 46 & 1245.87 & 25.35 & 41.52 \\ 
$\ldots$& $\ldots$& $\ldots$& $\ldots$& 47 & 1271.22 & 25.13 & 41.56 \\ 
$\ldots$& $\ldots$& $\ldots$& $\ldots$& 48 & 1296.35 & 25.98 & 41.58 \\    
\hline
\end{tabular*}
\end{table*}
\end{center}

On  the basis  of our  results,  we claim  that the  treatment of  the
chemical profile  at the chemical  transitions is a key  ingredient in
the computation of the g-spectrum. This is particularly true regarding
the mode  trapping properties, which  are considerably altered  when a
physically sound  treatment of the chemical  evolution is incorporated
in such  calculations. This conclusion  is valid at least  for massive
hydrogen   envelopes   as   predicted   by   our   full   evolutionary
calculations. To get  a deeper insight into these  aspects, we examine
the  node  distribution of  the  eigenfunctions  (Figs.   13 and  14).
According to  Brassard et al.   (1992a), this is an  useful diagnostic
for mode trapping.  Here, a  mode is trapped above the hydrogen-helium
interface when its  eigenfunction $y_1$ has a node  just above of such
an interface, and the corresponding node in $y_2$ lies just below that
interface.  Note that this statement is clearly satisfied by our model
with  diffusive equilibrium  in  the trace  element approximation,  as
shown in  Fig.  13. In  this figure, the  vertical dotted line  at $r=
0.927  R_*$  indicate the  location  of  the  pronounced peak  in  the
Brunt-V\"ais\"al\"a  frequency  (see  Fig.  9).  However,  the  node
distribution   becomes  markedly  different   in  the   sequence  with
non-equilibrium diffusion  and does not  seem possible to find  a well
defined interface in that case (see Fig.  14).  Thus, it does not seem
to be  clear that  the above-mentioned trapping  rule can  be directly
applied to  this case. In fact, because  the hydrogen-helium interface
becomes very smooth in our models, the peak in the Brunt-V\"ais\"al\"a
frequency  is not  very  pronounced.  Accordingly,  the capability  of
mechanical resonance of our model turns out to be weaker.  This causes
the node distribution in the eigenfunctions to be quite different from
that corresponding to the diffusive equilibrium approach.

\section{Conclusions} \label{sec:conclu}

In this work  we have explored the pulsational  properties of detailed
evolutionary    models    recently    developed    by    Althaus    et
al. (2002). Attention has been focused on a ZZ Ceti model in the frame
of  linear, non-radial  oscillations in  the  adiabatic approximation.
White dwarf  cooling has been  computed in a self-consistent  way with
the evolution  of the chemical  abundances resulting from  the various
diffusion processes and nuclear burning. Element diffusion is based on
a multicomponent gas treatment; so, the trace element approximation is
avoided  in our  calculations.  In addition,  the evolutionary  stages
prior  to   the  white  dwarf  formation  have   been  considered.  In
particular,  element  diffusion  causes  near discontinuities  in  the
chemical profile at the start of the cooling branch to be considerably
smoothed out by the time the ZZ Ceti domain is reached.

An important aspect of this work has been to assess the role played by
the  internal  chemical stratification  of  these  new  models in  the
behaviour  of  the  eigenmodes,  and  the expectations  for  the  full
g-spectrum  of  periods.  We  have  analyzed  mainly  the mode  weight
functions, which show  the regions of the star  that mostly contribute
to the period  formation.  Our study suggests the  existence of a much
wider diversity of eigenmodes for  periods shorter than $\approx 500 -
600$  s than found  in previous  works (Brassard  et al.  1992ab).  An
important  finding  of this  study  is  the  effect of  time-dependent
element diffusion on the mode  trapping properties in DA white dwarfs.
We conclude  that for periods longer  than $\approx 500 -  600$ s {\it
all} of  the modes seem to  be partially trapped  in the hydrogen-rich
envelope of  the star.   This conclusion, based  on the fact  that the
weight  functions of  these modes  show low  amplitudes  ($WF \lesssim
0.1$)  below the  hydrogen-helium transition  (even lower  as compared
with the  case of a  simulated chemically homogeneous  model), implies
that the  capability of  mode selection due  to mode  trapping effects
vanishes for high  periods when account is made  of white dwarf models
with diffusively  evolving stratifications.  This  conclusion is valid
at  least for  massive hydrogen  envelopes  as predicted  by our  full
evolutionary calculations.  This  behaviour is markedly different from
that  found in  other studies  based  on the  assumption of  diffusive
equilibrium in the trace element approximation.  This assumption leads
to  a  pronounced peak  in  the  Ledoux  term at  the  hydrogen-helium
interface, which is responsible for the trapping of modes in the outer
hydrogen-rich  layers.   We  have  verified this  fact  by  performing
additional  pulsation calculations on  a model  in which  the chemical
profile  at the  hydrogen-helium  transition is  given by  equilibrium
diffusion in the trace element approximation.  Finally, the prediction
of  both  diffusion  treatments  for  the  node  distribution  of  the
eigenfunctions has  been compared. We found that  node distribution at
the  hydrogen-helium  chemical  interface  is very  sensitive  to  the
treatment of the chemical profile at that interface.

On the basis of these new  results, we are forced to conclude that for
high  periods, trapping  mechanism  in massive  hydrogen envelopes  of
stratified DA  white dwarfs is not  an appropriate one  to explain the
fact  that all  the modes  expected  from theoretical  models are  not
observed in ZZ Ceti stars.  Interestingly, a weaker trapping effect on
the periodicities in DB white  dwarfs has also been reported (Gautschy
\& Althaus  2001).  We think that  the results presented  in this work
deserves further exploration from the point of view of a non-adiabatic
stability analysis. Work in this direction is in progress.

\begin{acknowledgements}
We warmly acknowledge Paul Bradley for providing us with his pulsational 
results about ZZ Ceti star models. We also acknowledge our referee, 
M. H. Montgomery, whose comments and suggestions strongly improved
the original version of this paper.
\end{acknowledgements}

\begin{figure*}
\centering
\includegraphics[width=450pt]{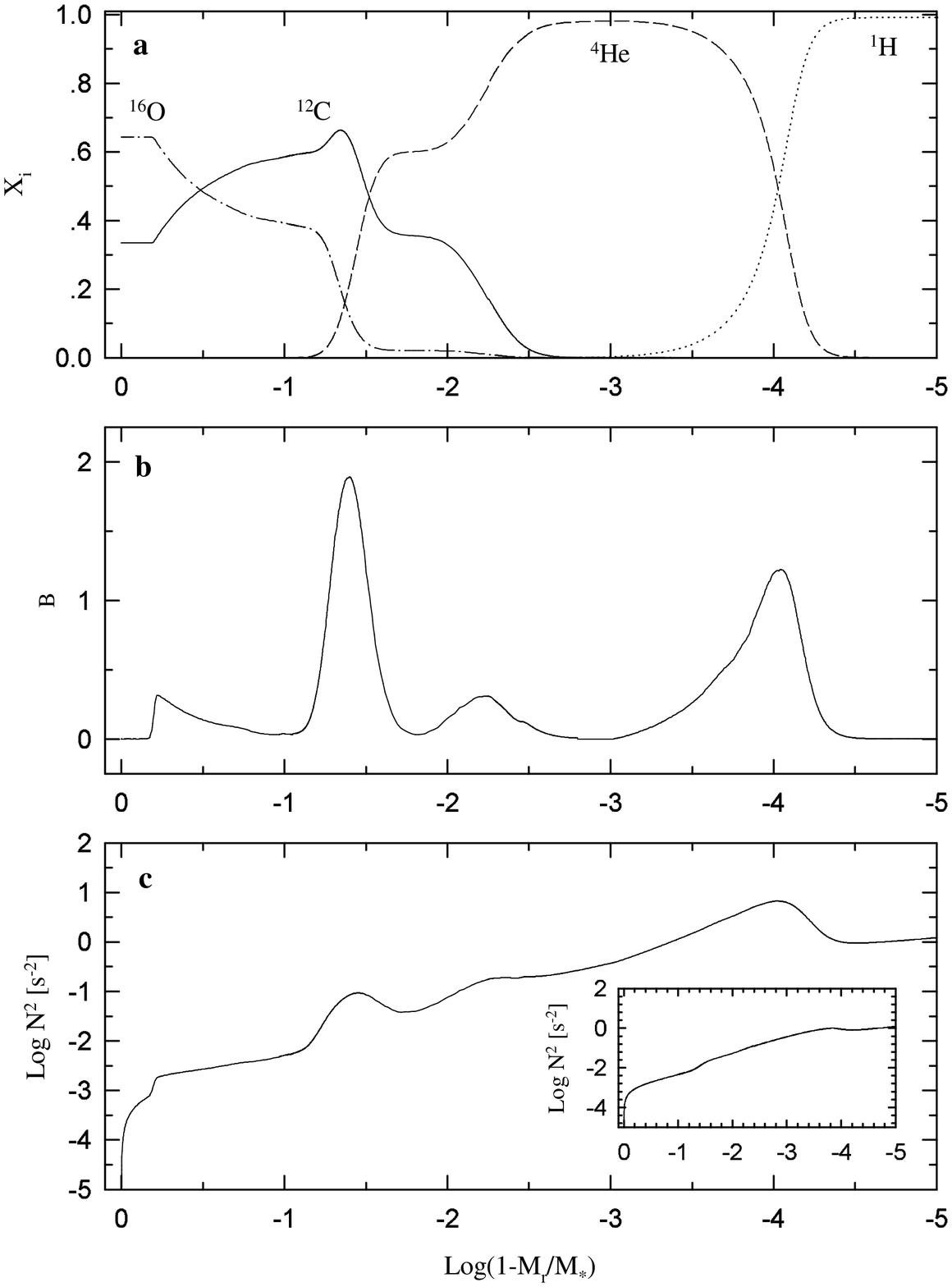}
\caption{Panel  {\bf a}:  The chemical  abundance distribution  of our
stellar model for hydrogen (dotted line), helium (dashed line), carbon
(solid line) and  oxygen (dot-dashed line). Panel {\bf  b}: the Ledoux
term,  $B$.   Panel {\bf  c}:  the logarithm  of  the  squared of  the
Brunt-V\"ais\"al\"a frequency  ($N^2$).  In  the inset of  this panel,
the  logarithm of  the  squared of  the Brunt-V\"ais\"al\"a  frequency
computed neglecting the term $B$  is depicted.  Note that the imprints
of the chemical transition zones in the
functional form  of $N^2$ are not completely eliminated,  in particular 
at  the helium-carbon-oxygen
and hydrogen-helium  interfaces.  The stellar mass of  the white dwarf
model  is 0.563  ${\rm M_{\odot}}$  and the  effective  temperature is
$\approx$  12000K. Quantities  are shown  in terms  of the  outer mass
fraction.}
\end{figure*}

\begin{figure*}
\centering
\includegraphics[width=450pt]{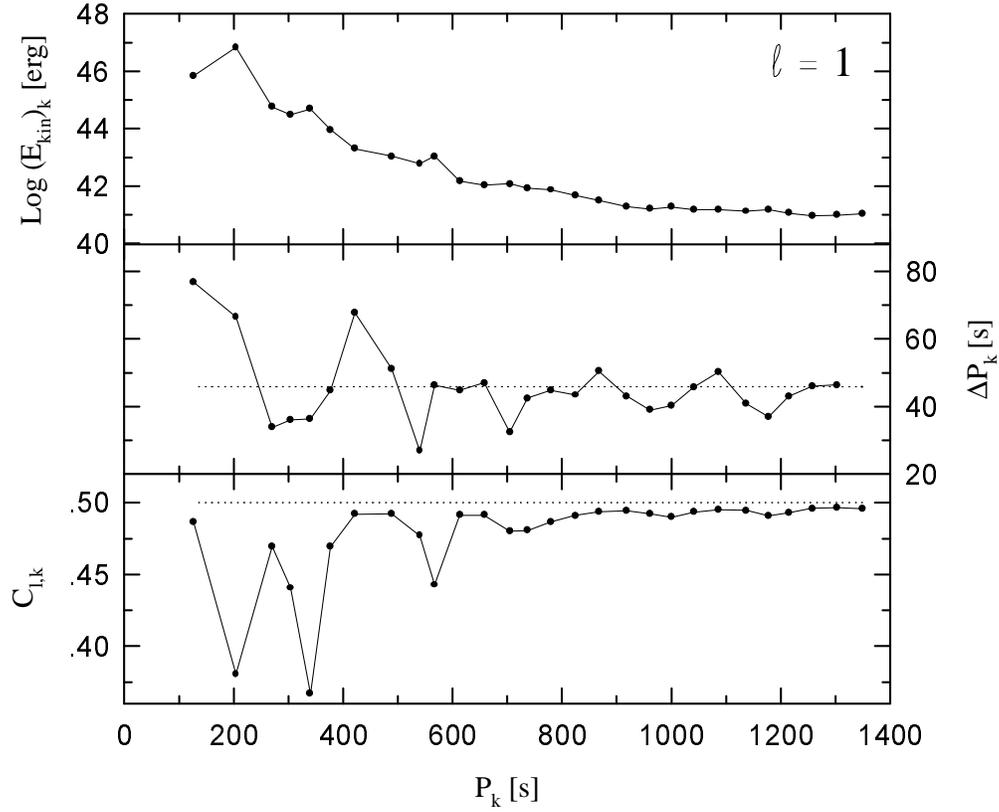}
\caption{The  logarithm  of the  oscillation  kinetic energy,  forward
period spacing and first  order rotation splitting coefficient (upper,
middle and lower  panels, respectively) for modes with  $\ell= 1$ as a
function  of  computed  periods.   Dotted lines  show  the  asymptotic
behaviour for  $\Delta P_k$ and $C_{\ell,k}$.  The  values of $(E_{\rm
kin})_k$ correspond to the normalization $y_1= \delta r / r= 1$ at $r=
R_*$.}
\end{figure*}

\begin{figure*}
\centering
\includegraphics[width=450pt]{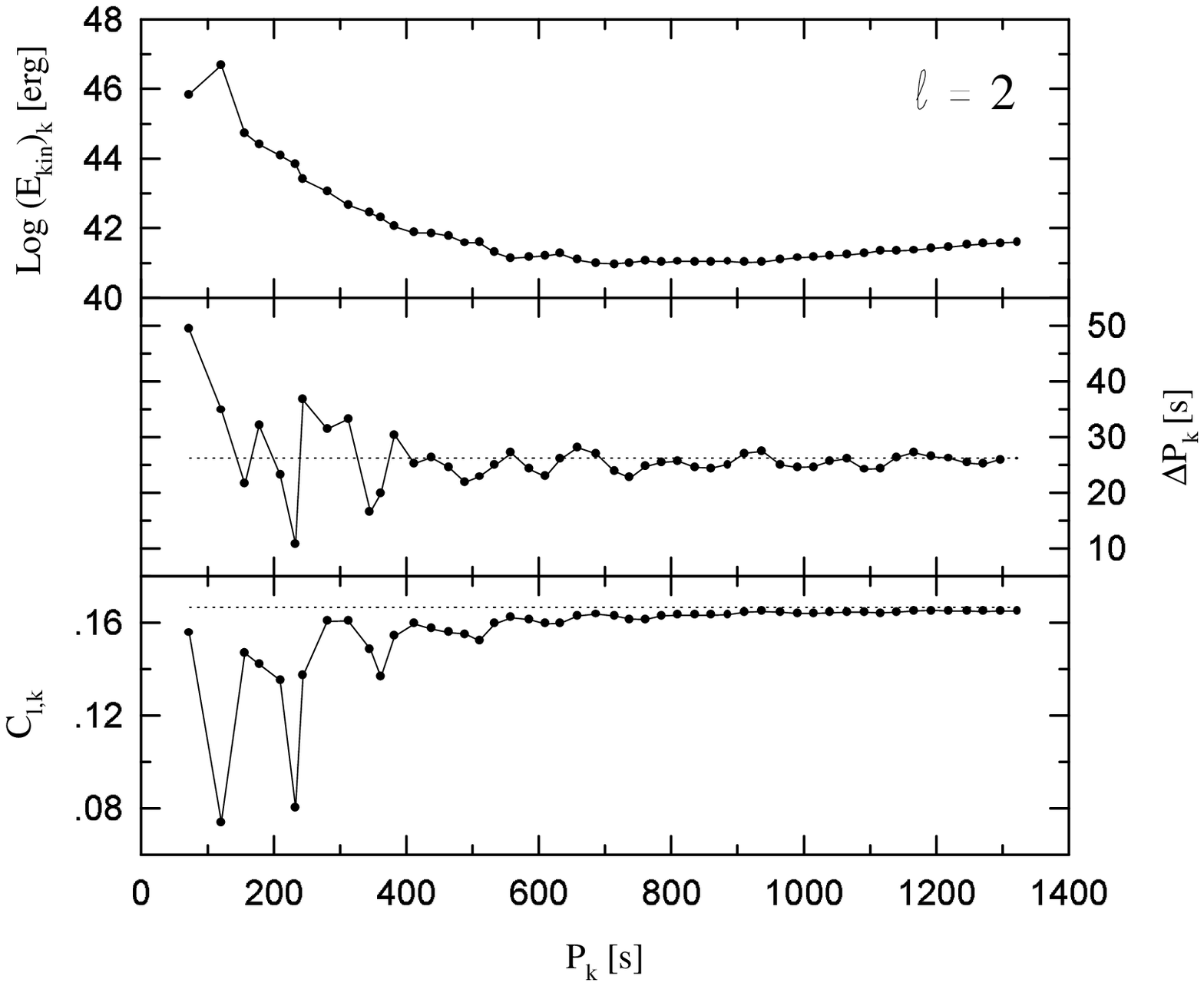}
\caption{Same as Fig. 2, but for $\ell=  2$.}
\end{figure*}

\begin{figure*}
\centering
\includegraphics[width=450pt]{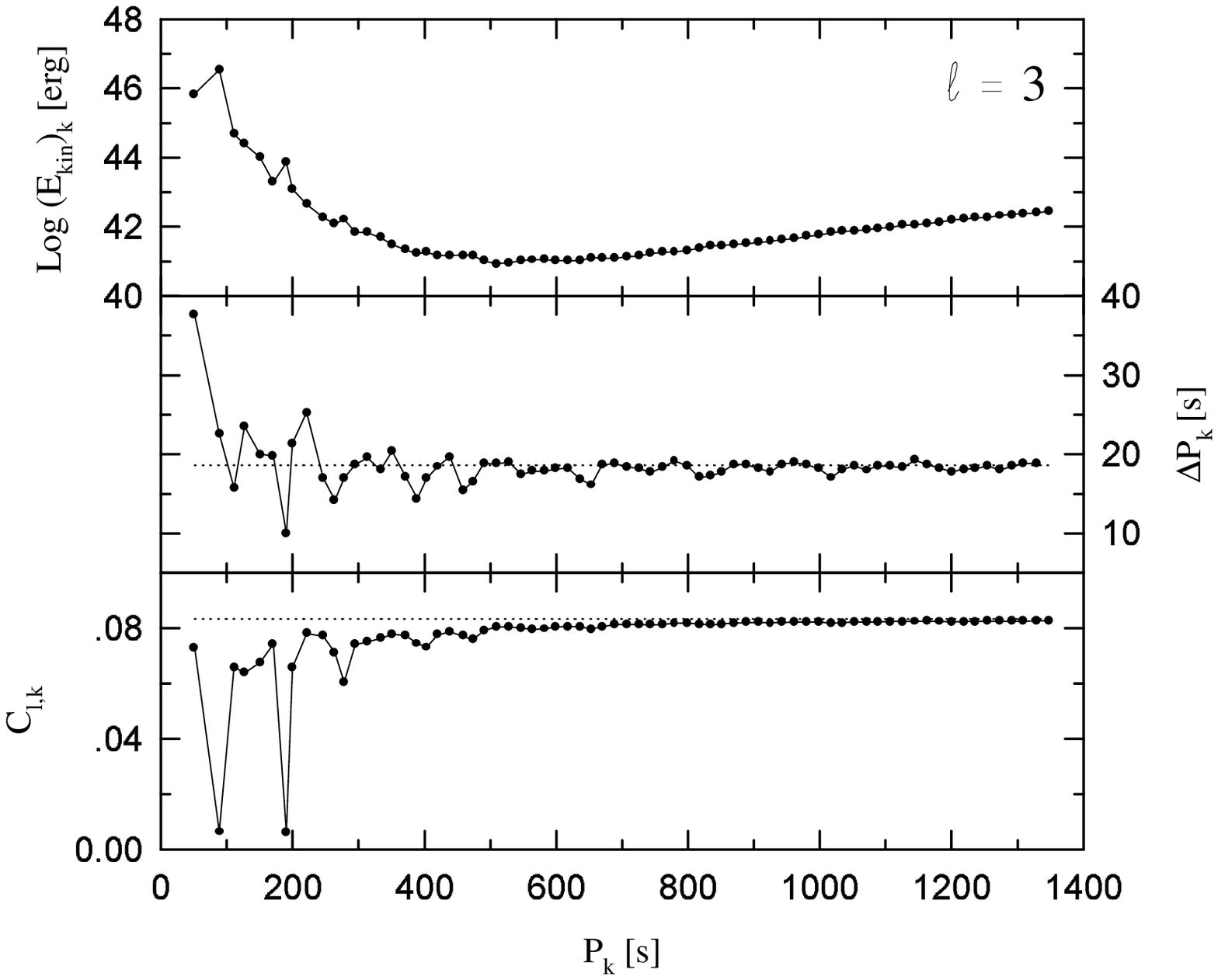}
\caption{Same as Fig. 2, but for $\ell=  3$.}
\end{figure*}

\begin{figure*}
\centering
\includegraphics[width=450pt]{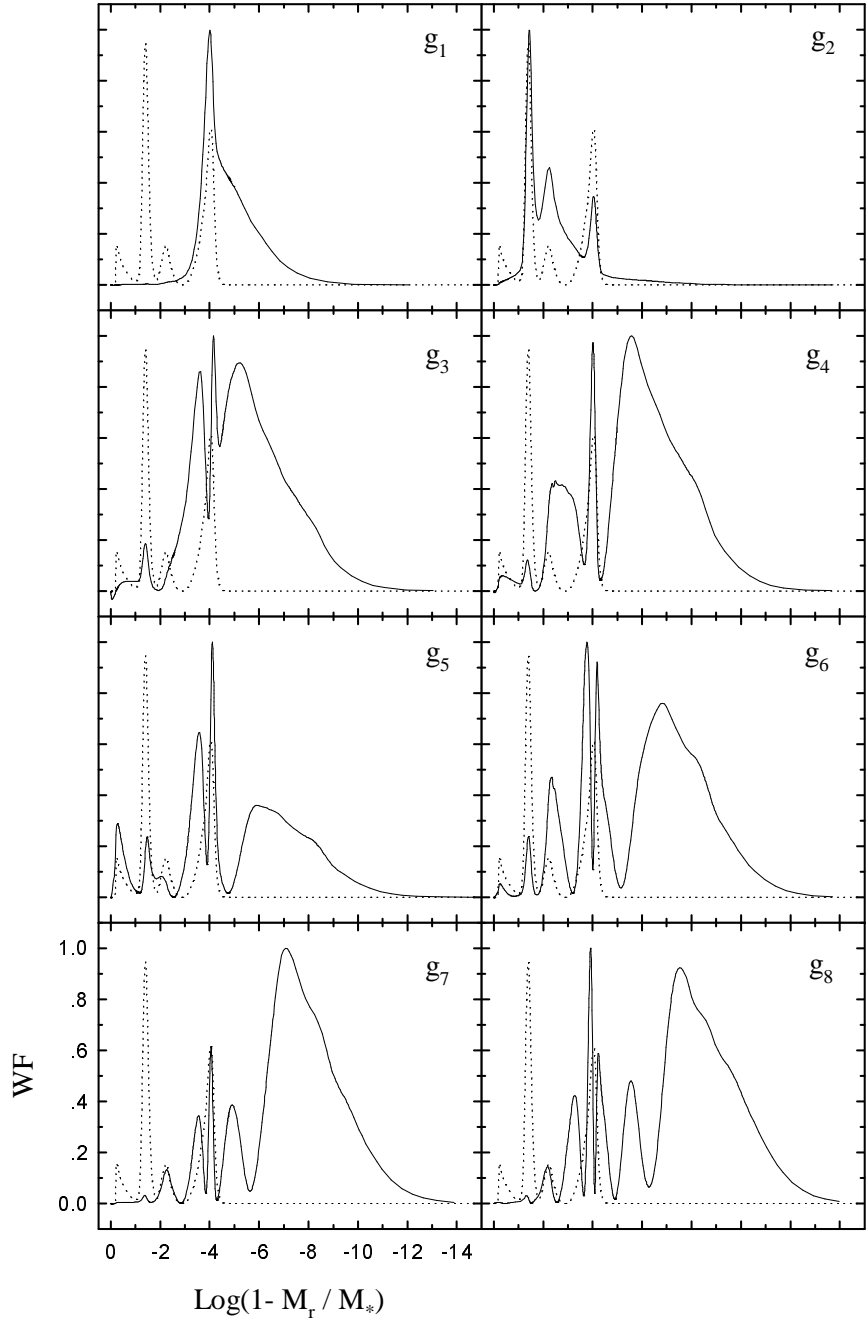}
\caption{The normalized weight function  (solid lines) in terms of the
outer mass fraction, for modes  g$_{1}$ to g$_{8}$ with $\ell= 1$.  In
the interests of comparison, dotted lines depict the run of the Ledoux
term (arbitrary units).}
\end{figure*}

\begin{figure*}
\centering
\includegraphics[width=450pt]{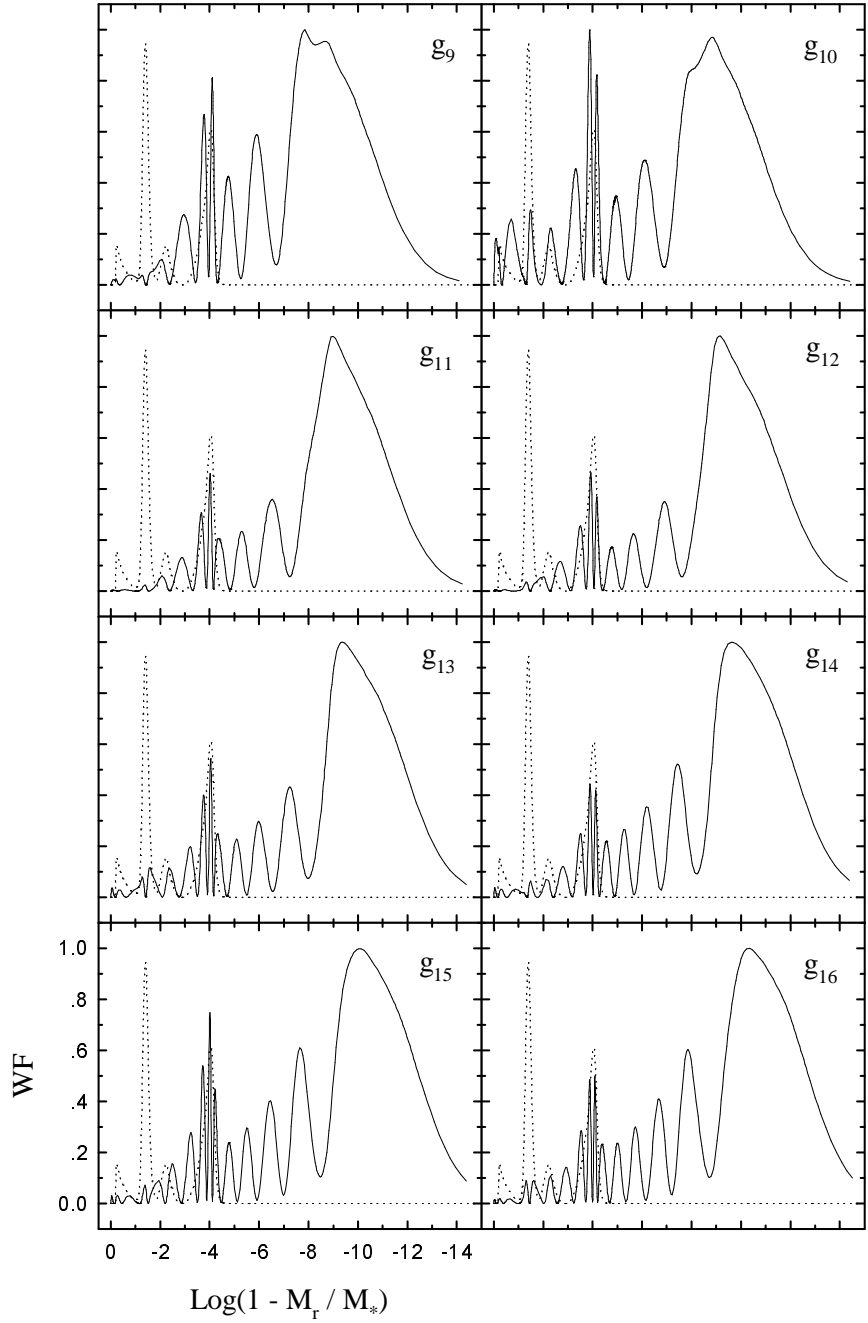}
\caption{Same as Fig. 5, but for modes g$_{9}$ to g$_{16}$.}
\end{figure*}

\begin{figure*}
\centering
\includegraphics[width=450pt]{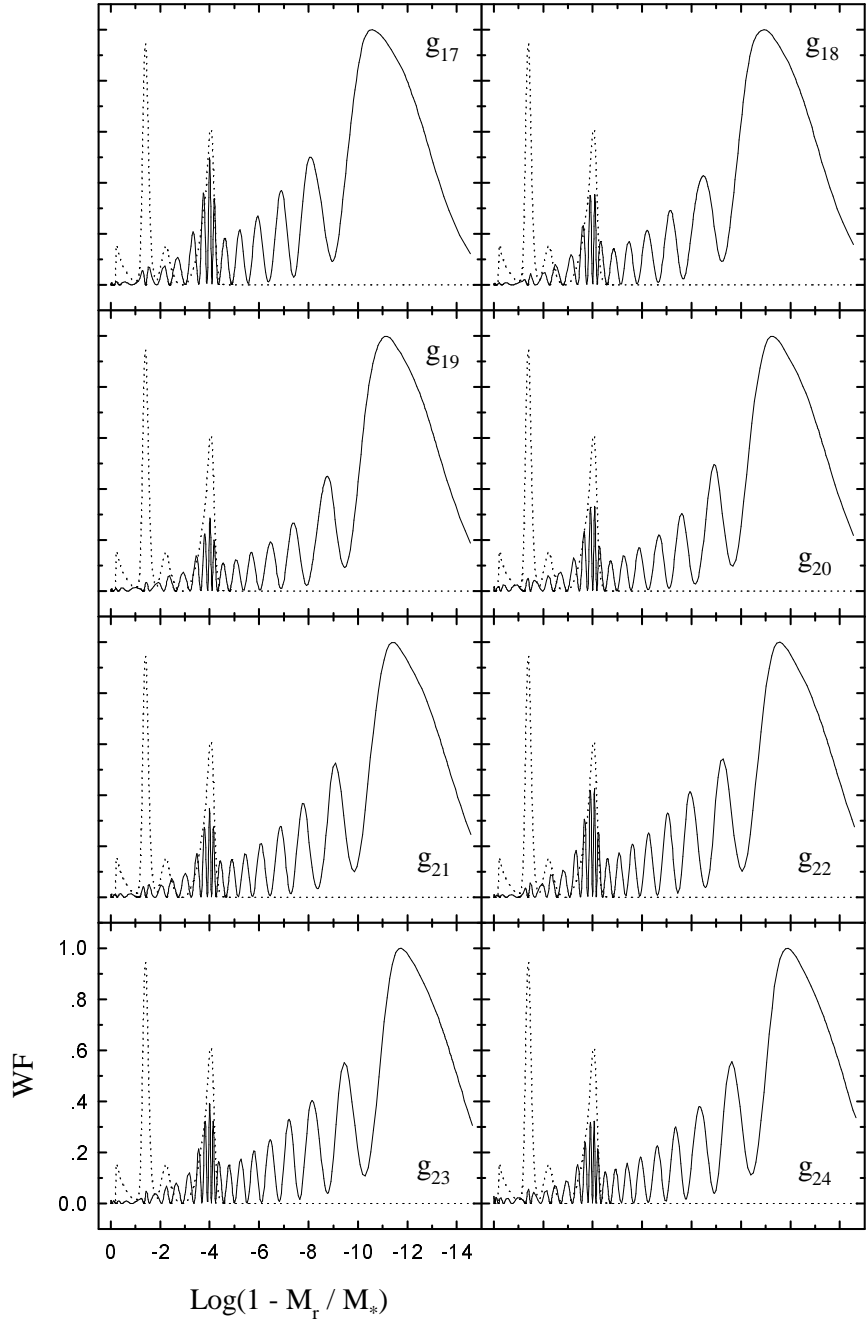} 
\caption{Same as Fig. 5, but for modes g$_{17}$ to g$_{24}$.}
\end{figure*}

\begin{figure*}
\centering
\includegraphics[width=450pt]{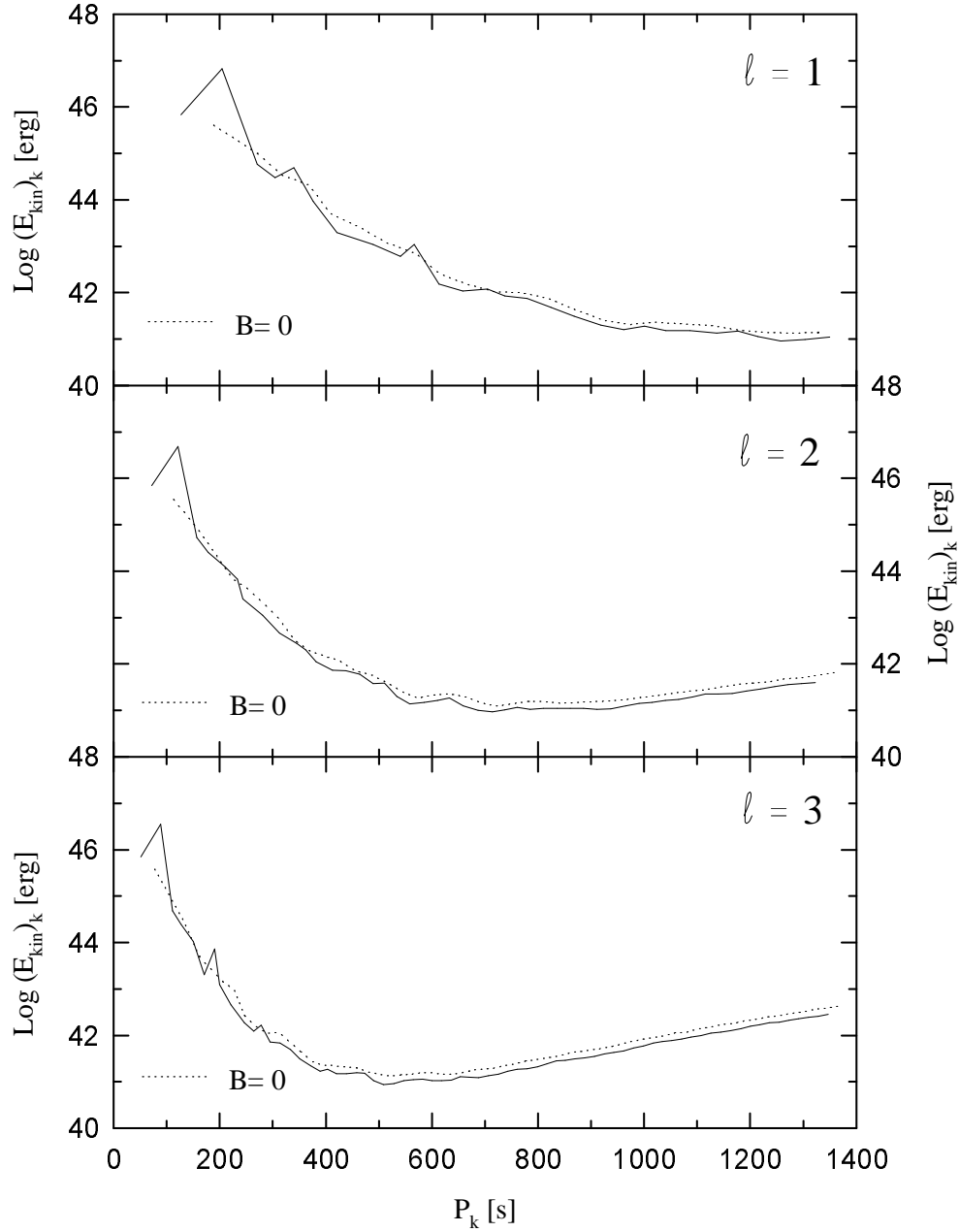}
\caption{The  logarithm of  the oscillation  kinetic energy  for modes
with  $\ell= 1$,  $\ell= 2$  and $\ell=  3$ (upper,  middle  and lower
panel,  respectively),  as a  function  of  computed  periods. In  the
interests  of clarity,  symbols corresponding  to eigenmodes  have been
omitted.   Solid lines correspond  to our  template model,  and dotted
lines correspond to the ``homogeneous'' model in which $B=$ 0.}
\end{figure*}

\begin{figure*}
\centering
\includegraphics[width=450pt]{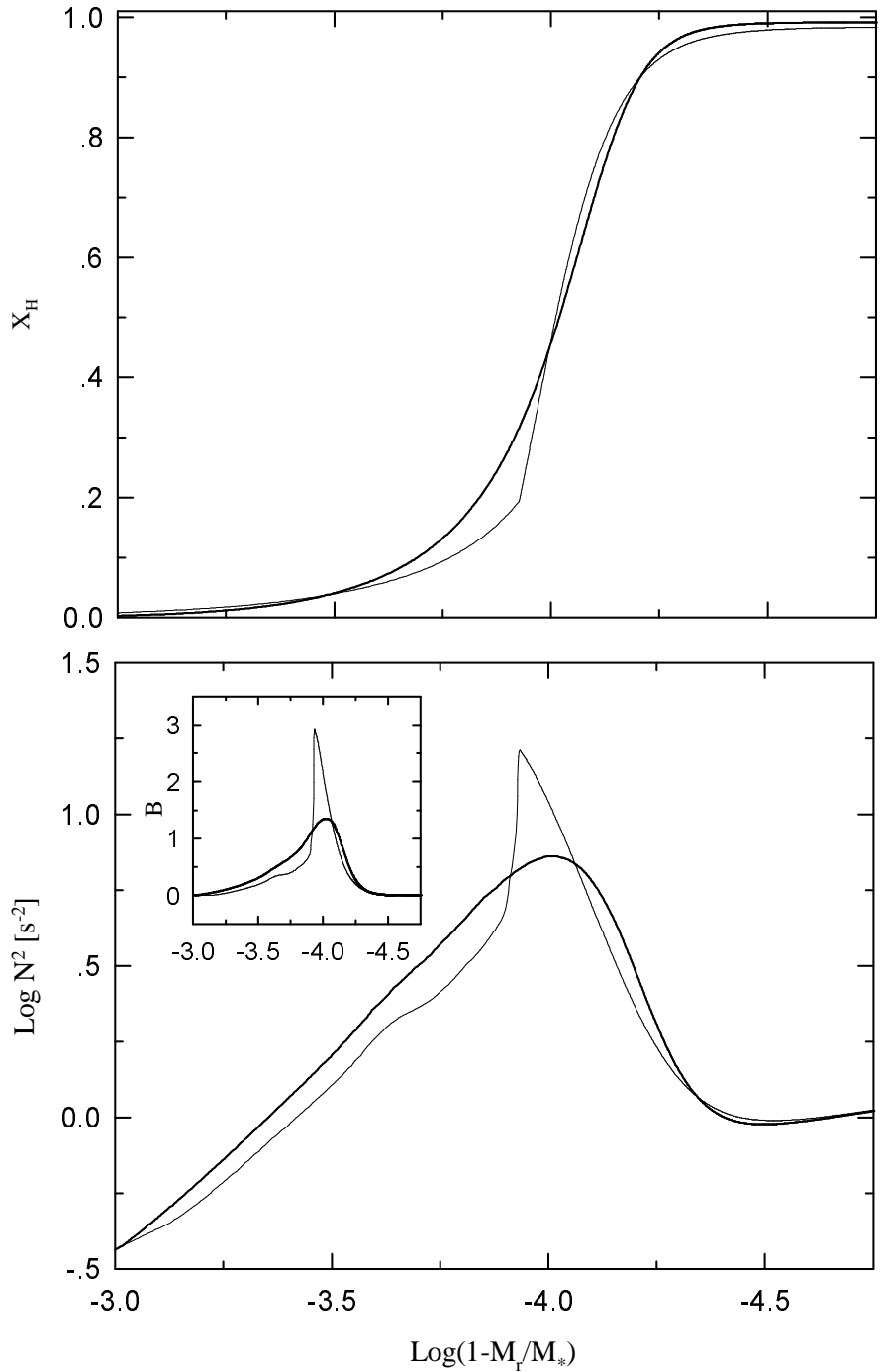}
\caption{Upper   panel:  hydrogen   abundance   distribution  at   the
hydrogen-helium interface as given by multicomponent, non-equilibrium 
diffusion (solid
line)  and diffusive  equilibrium in  the trace  element approximation
(thin   line).    Lower   panel:   the  logarithm   of   the   squared
Brunt-V\"ais\"al\"a  frequency  for the  both  treatment of  diffusion
mentioned  above.   The inset  shows  the  prediction  for the  Ledoux
term. For details, see text.}
\end{figure*}

\begin{figure*}
\centering
\includegraphics[width=450pt]{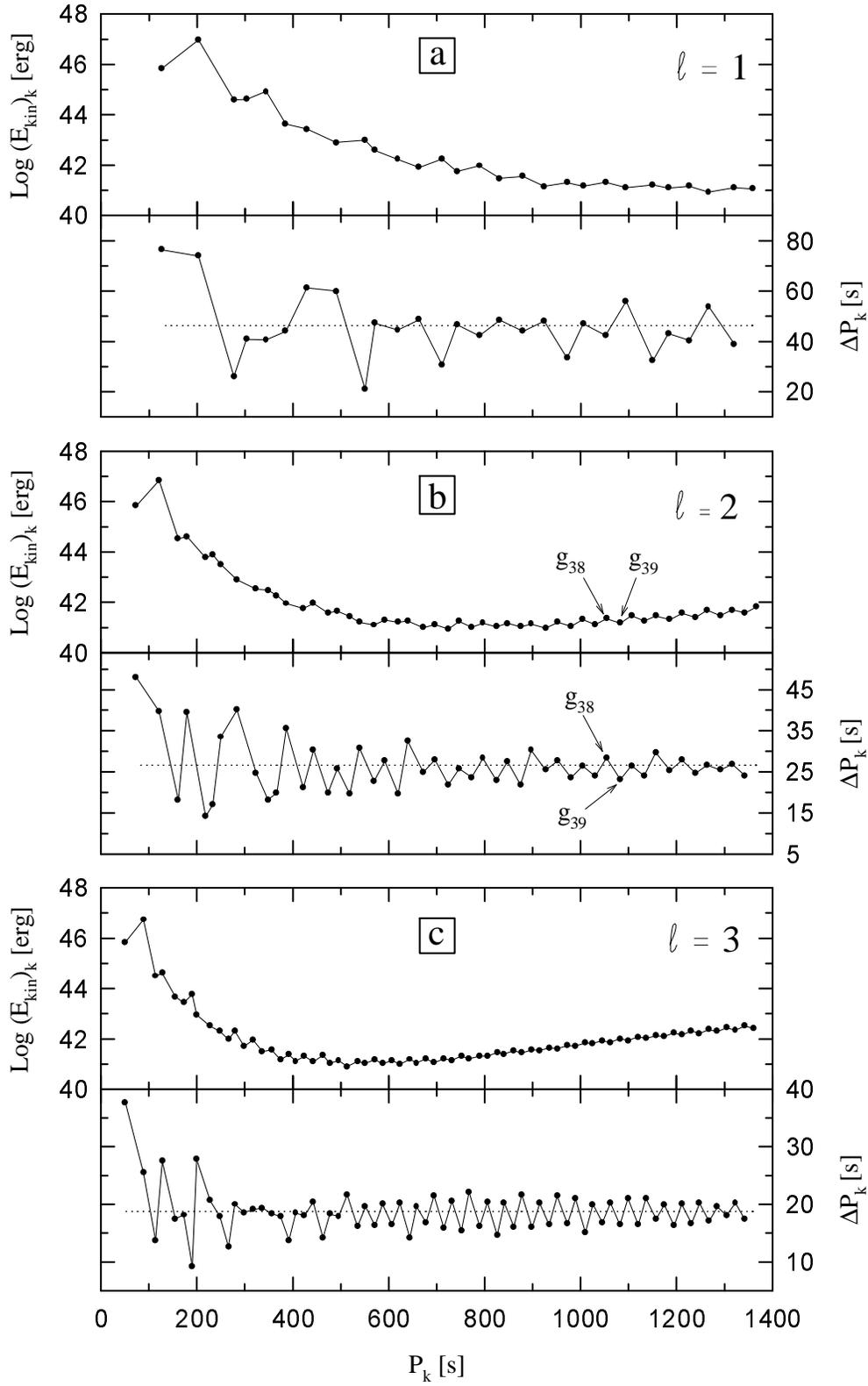}
\caption{The logarithm of the oscillation kinetic energy (upper panel)
and period spacing  (lower panel) for $\ell= 1$  (figure a), $\ell= 2$
(figure b) and $\ell= 3$ (figure  c) in terms of the computed periods,
for  the   case  of  diffusive   equilibrium  in  the   trace  element
approach. As found in  previous studies, this approximation gives rise
to  a kinetic  energy pattern  and spacing  of consecutive  periods in
which trapping signatures  are clearly noted.  This is  in contrast to
the prediction of time dependent  element diffusion given in Figs. 2
to 4.}
\end{figure*}

\begin{figure*}
\centering
\includegraphics[width=450pt]{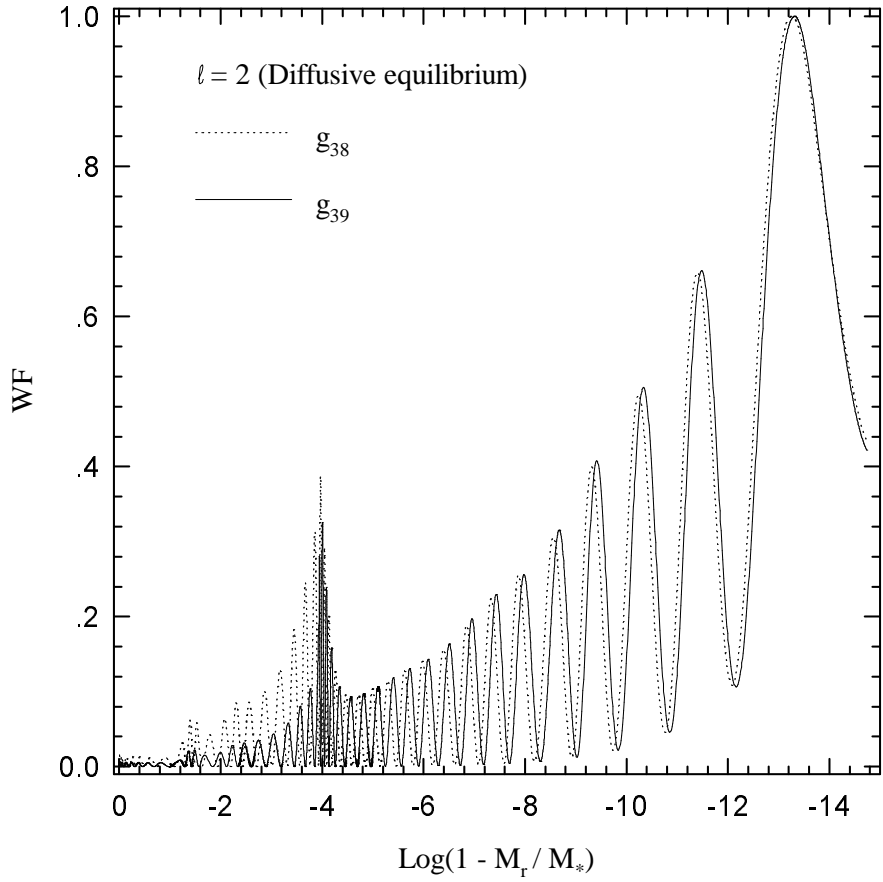}
\caption{The  normalized  weight   function  for  modes  g$_{38}$  and
g$_{39}$ with  $\ell= 2$ corresponding  to the stellar model  in which
the  hydrogen-helium  chemical  transition  has been  treated  in  the
diffusive  equilibrium and  trace element  approximation.  Note the  lower
amplitude of  $WF$ below the hydrogen-helium transition  for mode with
$k$= 39, which corresponds to a trapped one in the hydrogen envelope.}
\end{figure*}

\begin{figure*}
\centering
\includegraphics[width=450pt]{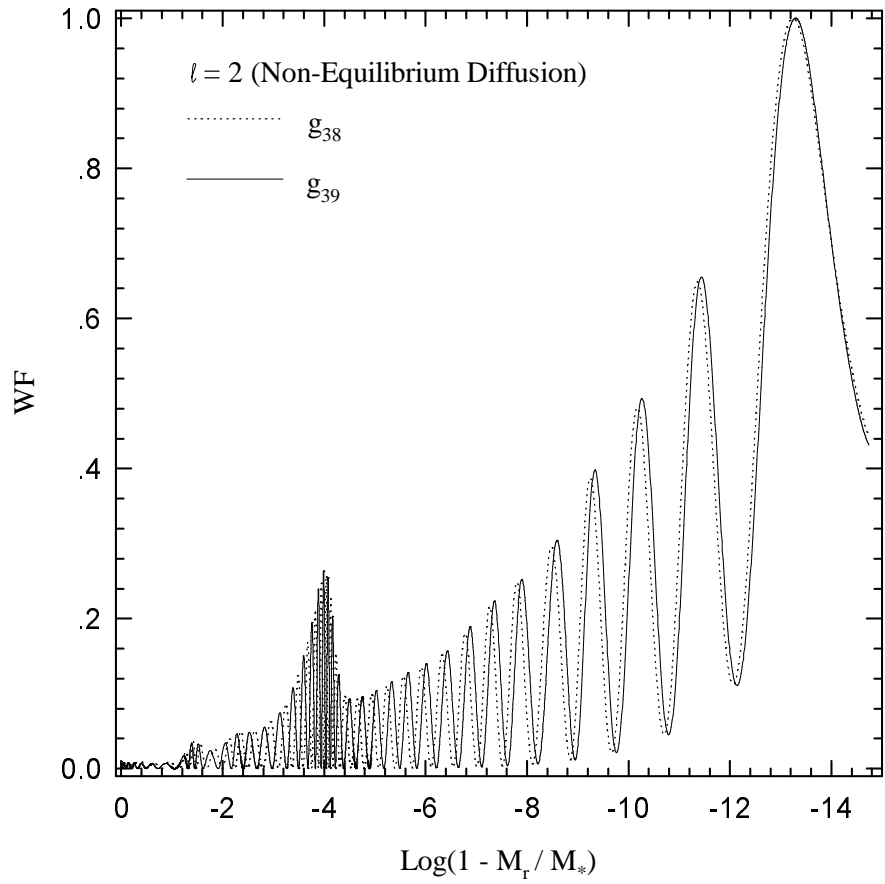} 
\caption{Same  of   Fig. 11,  but   for  the  case  in   which  the
hydrogen-helium  chemical transition has  been computed  assuming time
dependent element diffusion.}
\end{figure*}

\begin{figure*}
\centering
\includegraphics[width=450pt]{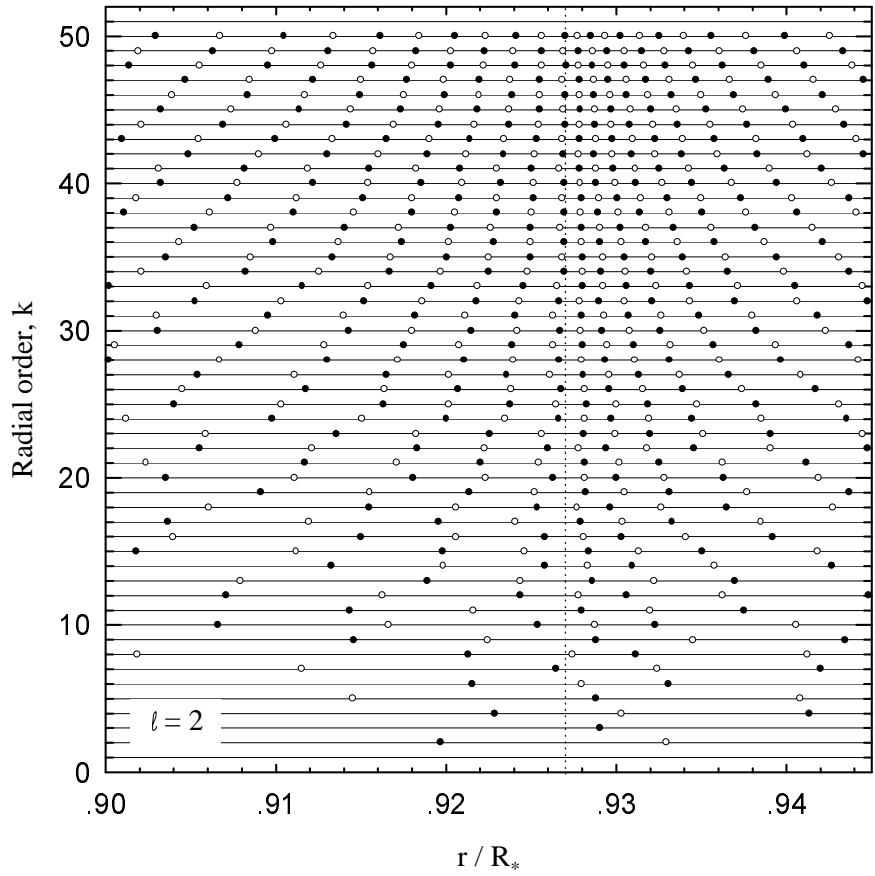}
\caption{The node  distribution of eigenfunctions  $y_1$ (filled dots)
and $y_2$  (empty dots) at  the hydrogen-helium transition  region for
modes  with   $\ell=  2$,  according  to   the  diffusive  equilibrium
prediction in the trace element approximation.}
\end{figure*}

\begin{figure*}
\centering
\includegraphics[width=450pt]{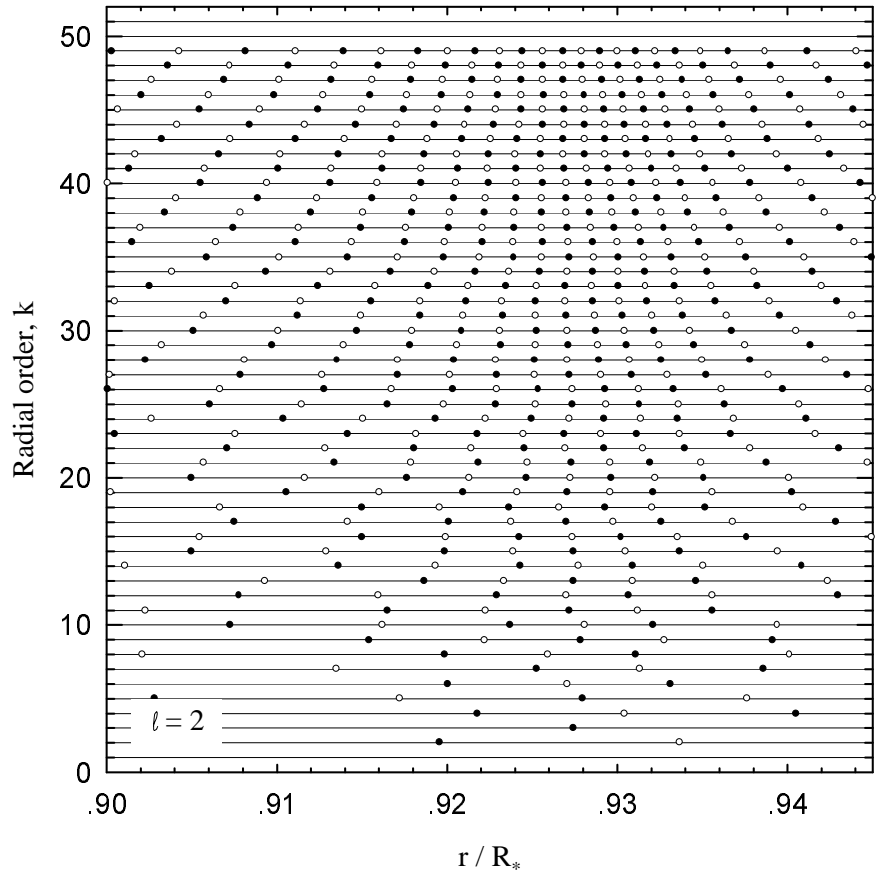}
\caption{Same  as Fig.  13, but  for the  case of  chemical profiles
resulting  from  time  dependent  element  diffusion.   See  text  for
details.}
\end{figure*}

\label{lastpage}

\end{document}